\documentclass[final]{siamltex}
\usepackage{latexsym,amssymb,lastpage,graphicx,amsfonts}
\usepackage{mathptmx,bm,amsmath,subfigure,color,dcolumn}

\newcommand{\heaviside}{\Theta}
\newcommand{\non}{\nonumber}
\newcommand{\beqa}{\begin{eqnarray}}
\newcommand{\eeqa}{\end{eqnarray}}
\newcommand{\beq}{\begin{equation}}
\newcommand{\eeq}{\end{equation}}
\newcommand{\dx}{\mathrm{d}x}
\newcommand{\dt}{\mathrm{d}t}
\newcommand{\pdhfrac}[2]{\mathchoice{\frac{#1}{#2}}{#1/#2}{#1/#2}{#1/#2}}
\newcommand{\fdd}[2]{\pdhfrac{\mathrm{d}#1}{\mathrm{d}#2}}
\newcommand{\pd}[2]{\pdhfrac{{\partial}#1}{{\partial}#2}}
\newcommand{\Sectionword}{Section{} }

\title{From Brownian Dynamics to Markov Chain: \\ an Ion Channel Example}

\author{Wan Chen$^1$\!\!\! \and \;
        Radek Erban$^{1,2}$\and \;
	S.~Jonathan Chapman$^{1,3}$}
	
\begin{document}

\maketitle

\footnotetext[1]{Mathematical Institute, 
	University of Oxford, 24-29 St Giles', Oxford OX1 3LB, United Kingdom}
\footnotetext[2]{E-mail: erban@maths.ox.ac.uk}
\footnotetext[3]{E-mail: chapman@maths.ox.ac.uk}

\begin{abstract}
A discrete rate theory for general multi-ion channels is presented, in which
the continuous dynamics of ion diffusion is reduced to transitions
between Markovian discrete states.
In an open channel, the ion permeation  process involves three types of
events: an ion entering the channel,
an ion escaping from the channel, or an ion hopping between different energy
minima in the channel.
The continuous dynamics leads to a hierarchy of Fokker-Planck
equations, indexed by channel occupancy. From these the mean escape times
and splitting probabilities (denoting from which
side an ion has escaped) can be calculated. By equating these with the
corresponding  expressions from the Markov model the Markovian transition
rates can be determined.
The theory is illustrated with a two-ion one-well channel.
The stationary probability of states is compared with that from
both Brownian dynamics simulation and
the hierarchical Fokker-Planck equations. The conductivity of the channel
is also studied, and the optimal geometry maximizing ion flux is computed.
\end{abstract}


\section{Introduction}

The membrane of a eukaryotic cell is mainly composed of a lipid bilayer, 
which is impermeable to water solvated ions \cite{Alberts:2007:MBC}.
Ion channels are nanopores formed by transmembrane proteins; they
allow ions to flow through and act as biological valves
connecting the intracellular with extracellular domains.
Ion channels are the main mechanism by which cells control the
intracellular concentration of chemical species, as well as the
potential gradient across the membrane.
As such they play important roles in maintaining various functions
of plant, animal and human cells.

There are two main features which distinguish ion channels  from other
nanoscale porous media. Firstly, they may be
selective, distinguishing between the charge and size
of ions; for example, the potassium K$^+$
channel conducts potassium ions at a rate $10^4$ times faster than
it  does sodium ions \cite{Doyle:1998:SPC}.
Secondly, their conformations may change between open and closed states
in response to an external stimulus such as a voltage gradient,
ligand binding, or pH value.

The molecular structure of many ion channels has been revealed by
X-ray crystallography in recent decades, which provides insight into
their features and function.
For example, the potassium K$^+$ channel  is composed of four identical
subunits which create a {\em cavity} connecting the cell interior to a
{\em selectivity filter}  at the outer end of the pore \cite{Doyle:1998:SPC}.
The narrow selectivity filter is only $12$ \AA \, long
and about $3$ \AA\, wide, which forces potassium ions with Pauling
radius $1.33$ \AA \,to shed their hydrating waters to enter and pass
in a single-file fashion. The oxygen atoms of four carbonyl
groups form four rings around the selectivity filter, which generate
local minima called {\em binding sites} in the overall energy
landscape to coordinate the dehydrated ions.

Mathematical models for ion channels include molecular dynamics
(MD), Brownian dynamics (BD) and continuum theory
(Poisson-Nernst-Planck equations) in descending order
of resolution \cite{Cooper:1985:TIT,Cooper:1988:DTD,Levitt:1999:MIC}. Molecular dynamics
provides the most detailed description by mimicking the motions and
interactions of all atoms (from membrane proteins to free ions and
even individual water
molecules) at the molecular level
\cite{Berneche:2001:EIC,Pongprayoon:2009:SAT,Jensen:2010:PCH}. Since the relaxation
of water molecules happens at
the fastest timescale of $1$ fs, the time step of an MD simulation has to
be very small, and one needs to evolve a system of thousands of
particles up to of the order of $0.1$ ms to observe ion conduction.
Such a simulation is obviously computationally intensive,
but much shorter simulations (of the order of $10$ ps)
can be used to obtain information about the local potential energy and the
effective diffusion coefficient of ions, which can then be fed into BD
simulations.

Brownian dynamics
\cite{vanGunsteren:1982:ABD,Corry:2000:TCT,Moy:2000:TCT,Cheng:2007:MFG} is a more
coarse-grained simulation which replaces the solvent molecules (water)
as a continuum, and represent their influence by a dialectric constant
and stochastic forcing.
The fluctuations of membrane proteins are ignored and the channel is
approximated by a solid boundary. Because the dynamics of water and
proteins are no longer
included, a relatively long time step can be used,  which greatly
reduces the computational cost. In this paper, we focus on this level
of resolution, and introduce a discrete rate theory that is based on
observations from BD.

The continuum model
\cite{Chen:1997:POC,Schuss:2001:DPN,Nadler:2004:IDC} calculates the potential
energy by a mean-field approximation of average ion positions, which
yields a Poisson equation,
and then formulates a Boltzmann equation (in equilibrium) or
a Nernst-Planck equation (in non-equilibrium) for the ion
concentration. These continuum partial differential equations (PDEs) 
can be solved efficiently; however
the individual ion-ion interaction is missing in this mean-field
assumption which then fails to predict some properties
(e.g. saturation). Comparisons of BD and continuum theories in different
channel configurations are presented in \cite{Corry:2000:TCT,Moy:2000:TCT}.

Recently several hybrid models combining MD and the theory of stochastic
processes have been proposed,
which are able to include molecular details and access long time scales
while keeping computational cost low.
One idea is to apply the Eyring rate theory to the ion permeation process
using the potential of mean force (PMF) calculated using MD.
This is based on the assumption that channels have some binding sites,
and ions pass through by a hopping mechanism: an ion
fluctuates around a certain site
before it obtains enough energy to overcome the energy barrier and
hops into the adjacent vacant site.
This ion hopping mechanism has been revealed by MD in channels with
binding sites \cite{Berneche:2001:EIC,Berneche:2003:MVI,Jensen:2010:PCH}.
In addition, the single file diffusion constraint imposed by the
narrowness of the channel assures that ions cannot cross each
other in the channel.
Therefore, the continuous dynamics of ion diffusion can be represented
by transitions between discrete Markovian states.

The Eyring rate theory was originally designed for chemical reactions
in the 1930s, with  transition rates  proportional to the
exponential of the energy barrier and distance between binding sites
\cite{Eyring:1935:ACA} (as shown in \cite{Cooper:1985:TIT}
this overestimates the physical barrier in the ion-crossing process).
A novel theory was proposed recently in \cite{Abad:2009:NRT} for a
one-dimensional channel with sawtooth-like PMF,
in which the transition rates are not approximated using the energy
barrier but are obtained as the product of total
escape rate from one binding site and the splitting probability
determining the relative chance of landing in each neighbouring site.
\cite{Abad:2009:NRT} showed that an optimal size of binding site maximizes
the ionic flux if the applied voltage exceeds a threshold.  They
assume the channel is occupied by at most one ion,
whereby the resulting system forms a single Markov chain,
and the rates can be solved explicitly.
In the multi-ion channel considered here,  ion-ion interactions as well
as the higher
dimension  of the energy landscape  mean that the complexity of the
rate theory is greatly increased.

In this paper, we present a general discrete rate theory for a multi-ion
channel, and compare it with BD.
The ion permeation process involves ion hopping, ion escaping and ion
entering.
For the purposes of this work
we assume ion entry rates are known and focus on calculating
the other rates in terms of the  mean escape time and splitting
probability.
Because of the complicated network between states the rates are more
intricately related to these quantities than in the single ion case.
Moreover,  since analytical solutions for the mean
escape time and splitting probability are not available,
these must be determined by solving the corresponding PDEs
numerically.
The theory is  illustrated by a two-ion channel with one binding site
and two ion sources. We show that, as with the  one-ion channel, there
exists an optimal shape for the external potential  that allows
a maximal flux.

The structure of this paper is as follows. In \Sectionword \ref{sec:BD},
we introduce a general theory for a multi-ion channel with a maximal
capacity of $N$ ions. We first present BD simulations
and formulate an equivalent cascade of hierarchical Fokker-Planck
equations for the probability distribution of ions. An illustrative
example of a $2$-ion channel is
discussed and the probability distribution from the histogram of BD
and the solution of the Fokker-Planck equation are compared.
Next, a discrete rate theory framework is presented in \Sectionword 
\ref{sec:rate} and the transition rates calculated. The
 $2$-ion channel is revisited in this
framework, and the result is compared with that from BD.
In \Sectionword \ref{sec:IV}, we apply
the theory to study the dependence of channel conduction on different
parameters such as the diffusion coefficient, ion entry rate and depth of
potential wells. In particular, we study the effect of the geometry of
the external
potential  in \Sectionword \ref{sec:geometry}. We conclude by
discussing the advantages and limitations of this method and possible 
applications and extensions in \Sectionword \ref{sec:conclusion}.

\section{Brownian dynamics} \label{sec:BD}

In this section, we present the theoretical framework of BD simulation. 
Since we are interested in studying the ion permeation process, which
occurs on  a 
time scale of $10^{-7}$ s,  and since conformational changes occur on
a timescale of $10^{-3}$ s, we assume that channel is always open and
does not change its 
conformation. 

Since the channel is very narrow and the ions pass through in single
file \cite{Jensen:2010:PCH},  we will suppose that the motion is 
one-dimensional, that is, the centres of
the ions will be constrained to lie along a line. The generalisation
to a fully three dimensional channel is algebraically complicated but
conceptually straightforward.
Since ions cannot pass each other in one dimension, we may neglect the
finite size of the ions and model them as point particles with
charge.

We define the maximal capacity of a channel to be $N$, so that it can
hold up to $N$ ions at one time.
We denote the number of binding sites in the channel by $M$. 
The parameters $N$ and $M$ vary among different channels; for example,
 a germicidal A channel has two
binding sites ($M=2$) and  single-ion occupation dominates (so that
$N=1$, or perhaps $N=2$ to allow for a knock-on effect) 
\cite{Abad:2009:NRT,Procopio:1979:ITF}.  

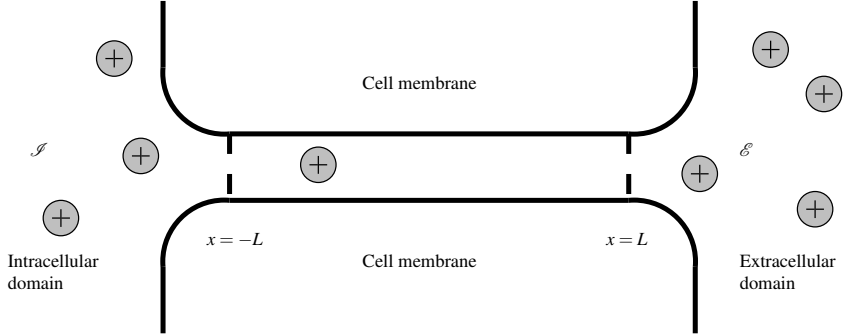
\begin{figure}[t]
\centering
{
\begin{picture}(0,0)%
\includegraphics{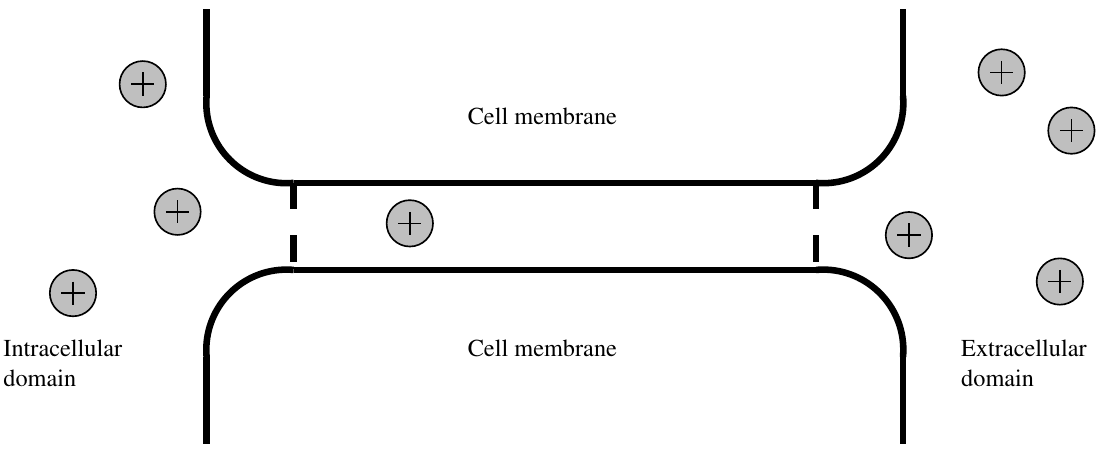}%
\end{picture}%
\setlength{\unitlength}{4144sp}%
\begingroup\makeatletter\ifx\SetFigFont\undefined%
\gdef\SetFigFont#1#2#3#4#5{%
  \reset@font\fontsize{#1}{#2pt}%
  \fontfamily{#3}\fontseries{#4}\fontshape{#5}%
  \selectfont}%
\fi\endgroup%
\begin{picture}(5043,2056)(886,-2291)
\put(2095,-1727){\makebox(0,0)[lb]{\smash{{\SetFigFont{7}{8.4}{\rmdefault}{\mddefault}{\updefault}{\color[rgb]{0,0,0}$x=-L$}%
}}}}
\put(4483,-1727){\makebox(0,0)[lb]{\smash{{\SetFigFont{7}{8.4}{\rmdefault}{\mddefault}{\updefault}{\color[rgb]{0,0,0}$x=L$}%
}}}}
\put(1034,-1196){\makebox(0,0)[lb]{\smash{{\SetFigFont{7}{8.4}{\rmdefault}{\mddefault}{\updefault}{\color[rgb]{0,0,0}${\cal I}$}%
}}}}
\put(5280,-1196){\makebox(0,0)[lb]{\smash{{\SetFigFont{7}{8.4}{\rmdefault}{\mddefault}{\updefault}{\color[rgb]{0,0,0}${\cal E}$}%
}}}}
\end{picture}%
}
\caption{A schematic structure of a channel. $x=-L$ and $x=L$
are the (artificial) left and right boundaries connecting 
large reservoirs of electrolyte in and outside the cell
respectively. $\mathcal{I}$ and $\mathcal{E}$ represent the overall
intracellular and extracellular environments, respectively. } 
\label{fig:channel}
\end{figure}

At the ends of the channel the pore opens out into the 
intracellular and extracellular space.
A full model would include (probably continuum) models of these
spaces, which would then be joined (preferably matched in terms of
matched asymptotic expansions, but more likely patched) 
to the channel model. For our present
purposes we need 
to introduce (artificial) interfaces (i.e. points) at the left and
right ends of the 
channel such that an ion passing through these interfaces is taken to
have left the channel and passed into the external domains.
Without loss of generality, we suppose that the left interface
connecting the channel to the intracellular domain lies at $x=-L$, and
the right interface connecting the channel to the 
extracellular domain lies at $x=L$, as shown in Fig. \ref{fig:channel}.
Thus an absorbing
boundary condition is imposed at $x=-L$ and $x=L$.

In BD simulation of an ion channel the contribution of water molecules
to the motion of a solute ion  can be approximated by random
collisions and an average frictional force in the evolution equation of
the solute ion \cite{vanGunsteren:1982:ABD,Moy:2000:TCT}. 
The motion of a system of $k$ ions is given by the Langevin equation
\begin{equation}
\label{fullLangevin}
m_i \, dv_i  = - \gamma\,v_i \mbox{d} t +
f^k_i(x_1, \ldots,x_k) \dt + \gamma
\sqrt{2 \,D}\,   dW_i, \quad i=1, \ldots, k, 
\end{equation}
where $x_i(t)$  and  $v_i(t)$ are the location and
velocity of the $i^{\mbox{\scriptsize th}}$ ion respectively.  
There are three forces on the right hand side of
(\ref{fullLangevin}). The first term 
corresponds to the frictional force exerted on the ion by averaging
the 
effect of water molecules; $\gamma$ is the frictional drag
coefficient, which depends on the surrounding fluid environment. Here
we assume  
it to be uniform so that $\gamma$ is constant. The third term is the
stochastic force generated by the random collisions of water molecules;
$W_i$ is a Wiener process and
$D=k_B T/\gamma$ is the diffusion coefficient, where
$k_B$ is the Boltzmann constant and $T$ is the temperature. 
The second term $f^k_i(x_1, \ldots, {x}_k)$ is the overall
electric force on the $i^{\mbox{\scriptsize th}}$ ion, 
including interactions with all other $k-1$ ions in the channel,  
fixed charges in the protein, and external field across the
membrane. It depends on the locations of all ions, and can be obtained
(along with the diffusion coefficient)
from MD simulation.  

Note that a typical value of the diffusion coefficient in aqueous
solutions at room temperature is $D \sim 10^{-3} \mbox{mm}^2 \mbox{s}^{-1}$,  
so that the ratio $m_i/\gamma \sim 10^{-14} \mbox{s}^{-1}$.  Since we
usually take a time step   
$\Delta t > 10^{-12} \mbox{s}$ in the simulation, the system is in an
overdamped limit \cite{Cooper:1985:TIT}. 
We may thus approximate (\ref{fullLangevin}) by the overdamped 
Langevin equation
\begin{equation} 
\label{Langevin}
 {dx}_i  =  \frac{\,D}{\,k_B T }  {f}^k_i({x}_1,
 \ldots, {x}_k)\, \dt 
+ \sqrt{2 D }\, {dW}_i, \quad i=1, \ldots, k.
\end{equation}
The boundary conditions on (\ref{Langevin}) may be described as
\begin{enumerate}
 \item When the number of ions in the channel $k$ is less than its
   capacity $N$, new ions are generated at the left (respectively
   right) end 
at a rate $H_k$ (respectively $G_k$). In principle  $H_k$ and $G_k$  
 depend  on the current locations of the  $k$ ions in the
channel ${x}_1, \ldots, {x}_k$ as well as the
intracellular and extracellular environments  
$\mathcal{I}$ and $\mathcal{E}$.

Since we are in the overdamped limit we cannot simply place the
incoming ions at the ends of the channel: under Brownian motion they
would immediately cross the boundary and leave the channel
again. Instead we place them at a position within the channel given by
the positional distribution function $h({x}; {x}_1, \ldots, {x}_k)$
(or respectively 
$g({x}; {x}_1, \ldots, {x}_k)$). Note that $h$ and $g$ also depend  on
the positions of the existing ions. This is necessary since, for
example, an ion entering the channel from the left must lie to the
left of $x_1$, while an ion entering from the right must lie to the
right of $x_k$. Thus, at the very least, $h$ depends on $x_1$ while
$g$ depends on $x_k$.

The functions $h$ and $g$ should  be chosen to make the
join with the outer model as smooth as possible, as in 
\cite{Flegg:2012:TRM,Franz:2012:PGS}.
Here we simply assume that $h$ and $g$, and
the rates $H_k$ and $G_k$, are given.

\item When ${x}_i(t)<-L$ or ${x}_i(t)>L$  the $i^{\mbox{\scriptsize th}}$ 
ion is removed from the channel.  
\item If ${x}_i(t)> x_{i+1}(t)$ for some $i$ then $x_i$ and $x_{i+1}$
  are switched. This enforces the single-file nature of the channel by
  preventing an ion overtaking its neighbour. 
This  condition is unlikely to occur with ions in
  a channel due to the strong Coulomb repulsion, but may be necessary
  if we are interested in neutral molecules.
\end{enumerate}

\subsection{Hierarchical Fokker-Planck equations}

We denote by
$P_k({x}_1, \ldots, {x}_k, t)$ the probability density
function for the event that there are $k$ ions in the channel at positions 
${x}_1, \ldots, {x}_k$ at time $t$. Since the number of
ions in the channel may run from zero to the channel capacity $N$, we
have $N+1$ such probability density functions.
The probability of no ion in the channel (i.e. $k=0$) 
is denoted by $P_0(t)$, and is independent of the spatial variable. 
 We label the ions by the order of their locations, such that
${x}_i < {x}_j$ 
for $i <j$. Then the stochastic process     
\eqref{Langevin} is equivalent to the following hierarchical system of
Fokker-Planck equations:

\begin{figure}[t]
\centering
{
 \includegraphics[width=5.04in]{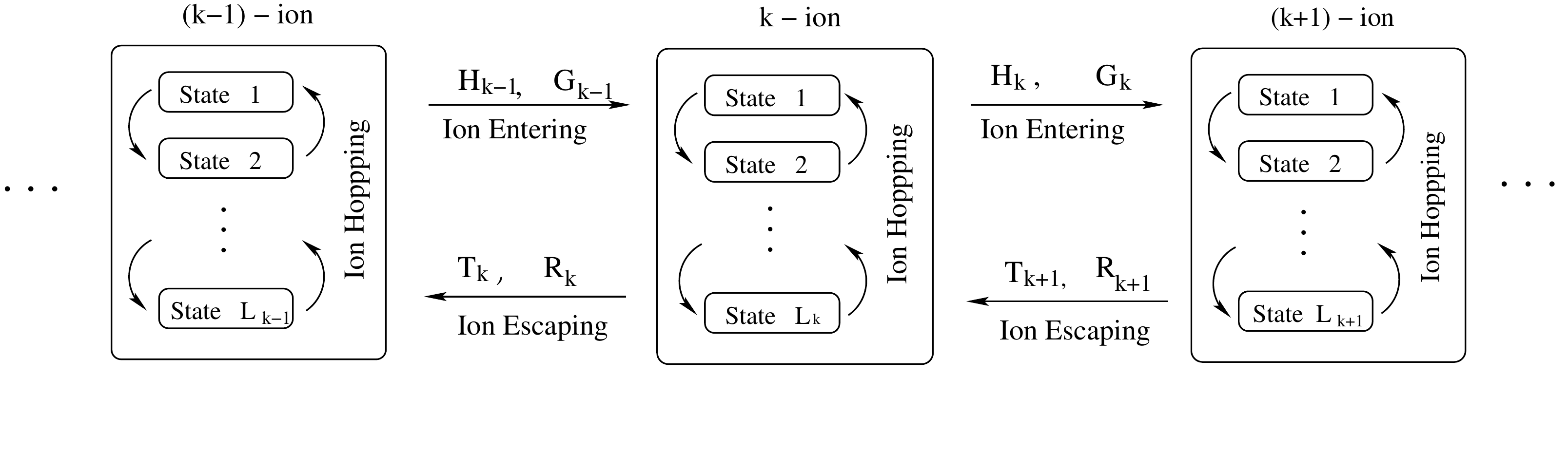}
}
\caption{Hierarchical Fokker-Planck Equations describe the conservation of
 ions in the channel. For  $k$-ion occupancy, the transitions to and from 
 $(k-1)$-ion and $(k+1)$-ion occupancy by ions entering and escaping  
are demonstrated, along with internal transitions between states.}
\label{fig:FP}
\end{figure}
 
\begin{subequations}
\begin{eqnarray}
\partial_t\, P_k({x}_1, \ldots, {x}_k,t) &=& D \,\nabla
\cdot \Big( \nabla P_k({x}_1, \ldots, {x}_k,t) 
 -  P_k({x}_1, \ldots, {x}_k,t) \, \frac{\,1}{\,k_B T}
 {\mathbf F}_k({x}_1, \ldots, {x}_k,t)
 \Big)\nonumber \\  
&&\mbox{}- \Big(H_k({x}_1, \ldots, {x}_k)
+  G_k({x}_1, \ldots, {x}_{k})
\Big)  P_k({x}_1, \ldots, {x}_k, t) \nonumber \\  
&&\mbox{}+  H_{k-1}({x}_2, \ldots,
{x}_{k})\,h({x}_1;x_2,\ldots,x_{k})P_{k-1}({x}_2, \ldots, 
{x}_{k}, t) \nonumber \\  
&& \mbox{}  +  G_{k-1}({x}_1, \ldots,
{x}_{k-1})\, g({x}_k;x_1,\ldots,x_{k-1}) 
P_{k-1}({x}_1, \ldots, {x}_{k-1}, t)\nonumber \\  
&& 
\mbox{}+ T_{k+1}( {x}_1, \ldots, {x}_{k}) +
R_{k+1}( {x}_1, \ldots, {x}_{k}) , \label{FP}
\end{eqnarray}
where $\nabla=(\partial_{{x}_1},
\ldots, \partial_{{x}_k})$, ${\mathbf F}_k=({f}^k_1,
\ldots, {f}^k_k) \in \mathbb{R}^{k}$, and  
\begin{eqnarray*}
T_{k+1}( {x}_2, \ldots, {x}_{k+1}) &=&   D \Big(
\frac{\partial P_{k+1}}{\partial x_1}  - 
P_{k+1} \, \frac{\,1}{k_B T} {f}^{k+1}_1 \Big)(-L,x_2,\ldots,x_{k+1}) , \\  
R_{k+1}( {x}_1, \ldots, {x}_{k})& =&  -D \Big( \frac{\partial
  P_{k+1}}{\partial x_{k+1}}    - 
P_{k+1} \, \frac{\,1}{k_B T} {f}^{k+1}_{k+1} \Big)(x_1,\ldots,x_{k},L)  , 
\end{eqnarray*}
where $k =0,\ldots,N$ and we use the convention that $P_{-1} =
P_{N+1} = 0$.
Since only one ion can
escape or enter at any one time, $P_k$ is  coupled only to the neighbouring
states $P_{k-1}$ and $P_{k+1}$. Note that $H_{N} = G_{N} = 0$, since
no ions can enter when the channel is fully occupied.

The first two terms (i.e. the first line)
on the right-hand side of (\ref{FP}) correspond to
ion diffusion and ion drift respectively, where the drift term
includes the external potential as well as ion-ion interactions. The
third term corresponds to a new ion entering the $k$-ion channel from
intracellular or extracellular solution; this term is negative since
such an event leads to a transition from a $k$-ion channel to a
$(k+1)$-ion channel. The fourth and fifth terms correspond to a new ion
entering a $(k-1)$-ion channel from the left and right respectively.
The sixth and seventh terms (i.e. the last line) of (\ref{FP}) 
correspond to ions leaving a $(k+1)$-ion
channel from the left and right respectively.

\noindent
The boundary conditions on (\ref{FP}) are 
\begin{eqnarray}
P_k(-L,x_2,\ldots,x_{k}) & = & 0,\\
P_k(x_1,\ldots,x_{k-1},L) & = & 0,
\end{eqnarray}
along with the no-flux condition on the interface $x_i = x_{i+1}$,
\begin{equation}
\lim_{x_i \rightarrow x_{i+1}} 
\left(\frac{\partial P_k}{\partial x_i} - P_k \frac{1}{k_B T} f_i^k
\right)
 = \lim_{x_{i} \rightarrow x_{i+1}}\left(  \frac{\partial P_k}{\partial
  x_{i+1}} - P_k \frac{1}{k_B T} f_{i+1}^k
\right)
\end{equation}
\end{subequations}
for $i = 1,\ldots,k-1$, which ensures that the ions are correctly labelled.
The reason we have had to write this as a limit is that the 
 inter-ion potential tends to infinity as $x_i \rightarrow x_{i+1}$, while  
$P_k$ tends to zero. A local analysis shows that we need $P_k$ to tend
to zero faster than $(x_{i+1}-x_i)^2$.

We note the following normalisation condition, which 
 holds at all times $t$,  
\begin{equation}
\label{sumP}
\sum_{k=0}^N \int_{\Gamma_k} P_k({x}_1, \ldots, {x}_k,t)\, d {x}_1
\cdots d {x}_k = 1,
\end{equation}
where $\Gamma_k$ is the available state space when there are $k$ ions
in the channel, namely $\Gamma_k = \{(x_1, \ldots,x_k): x_1<x_2<
\cdots < x_k\}$.
We will usually be interested in the steady state; in that case  
we solve the
coupled hierarchical Fokker-Planck 
equations for the stationary probability distribution 
$\widetilde{P}_k = \lim_{t \to \infty} P_k(t) $ for  
$k=0, 1, \ldots, N$. 

\subsection{An example with $N=2$} \label{sec:example}

We exemplify the  theory above with a simple channel that is
selective to cations with elementary charge $e = 1.6 \times
10^{-19}\,\mbox{C}$.  
The selectivity of this type of channel is generally caused by 
negative charged boundary proteins, 
which decrease the energy barrier imposed by the narrow structure and
assist the permeation of cations.  
For example, the oxygen atoms of four carbonyl groups in the
selectivity filter of the potassium channel can be modelled by putting
four negative partial charges equally 
spaced on a ring of radius $d$ that is perpendicular to the $x$-axis
\cite{Corry:2000:TCT}. 

We consider the simplest possible example of multi-ion channel 
with capacity $N=2$ and a single binding site $M=1$. 
The binding site is located  at the position $x=\xi$ and is a potential
well generated by a ring of fixed partial negative charges  a
distance $d$  from the channel axis.   
By Coulomb's law, the potential energy  $\Phi_1(x_1)$ seen by one
cation at $x_1$ with charge $e$ traversing through the channel is 
\begin{subequations}
\label{pot_formula}
\begin{equation}
\label{pot1}
\Phi_1(x_1) =   \frac{\,e}{\,k_B T } 
\Big( \frac{- k_e Z }{\sqrt{(x_1-\xi)^2+d^2}} + U x_1 \Big).
\end{equation}
where $k_e$ the Coulomb force constant and $Z$
the total fixed charge on the ring, and 
$U$ is the constant field, which imposes a potential difference $2 U
L$  across the channel $[-L, L]$. This potential difference  
is small compared to  the potential  well, 
and does not change the shape of the potential well but merely tilts 
it by a small angle. The force on the ion due to the potential is 
\[ 
f_1^1 = - k_B T \,\frac{{\rm d} \Phi_1}{{\rm d} x_1}
.\]

When there are two cations in the channel, at positions $x_1$ and $x_2$,
the overall potential energy  $\Phi_2(x_1,x_2)$, including the interaction 
between the two free ions, is
\begin{equation}
\Phi_2(x_1,x_2) =  \frac{\,e}{\,k_B T } \Big( \frac{- k_e Z
}{\sqrt{(x_1-\xi)^2+d^2}} + \frac{- k_e Z }{\sqrt{(x_2-\xi)^2+d^2}} +
\frac{k_e e}{|x_1-x_2|} + U (x_1 + x_2) \Big). 
\end{equation}
\end{subequations}
The forces on the two ions are then
\[
f^2_1 = - k_B T\,\frac{\partial \Phi_2}{\partial x_1},
\qquad
f^2_2 = - k_B T\,\frac{\partial \Phi_2}{\partial x_2}.
\]
Finally we need to specify the entry rates $H_k$ and $G_k$ and entry
distribution functions $h$ and $g$. 
We choose the simplest possible model for the entry distribution
function. We suppose that the ions entering from the left are all
placed at a position $x_-$ near the left-hand end of the channel,
while ions enterying from the right are placed at a position $x_+$
near the right-hand end of the channel, that is
\[ h(x) =   \delta(x -x_-) ,
\qquad
g(x) =   \delta(x -x_+) .
\]
 We have to be careful in
implementing this condition that we preserve the order of the ions in
the channel. We choose to do this as follows: if we are attempting to
place an ion at position $x_-$, and the position of the existing
 ion $x_1<x_-$, then we abandon the insertion of the new ion.
A similar procedure is implemented at the right-hand end.
In effect this means that the rate of entry is chosen to be zero
whenever the position $x_1$ of the existing ion is such that $x_1<x_-$
or $x_1>x_+$. (An alternative procedure would be to modify the
distribution functions $h$ and $g$ so that $h=0$ if $x_1<x_-$ and
$g=0$ if $x_1>x_+$, but this would mean altering them from the present
$\delta$-functions.) 

\begin{figure}[t]
\centering
\subfigure[ Histogram ]
{\label{fig:SSProb2}
 \includegraphics[width=2.35in]{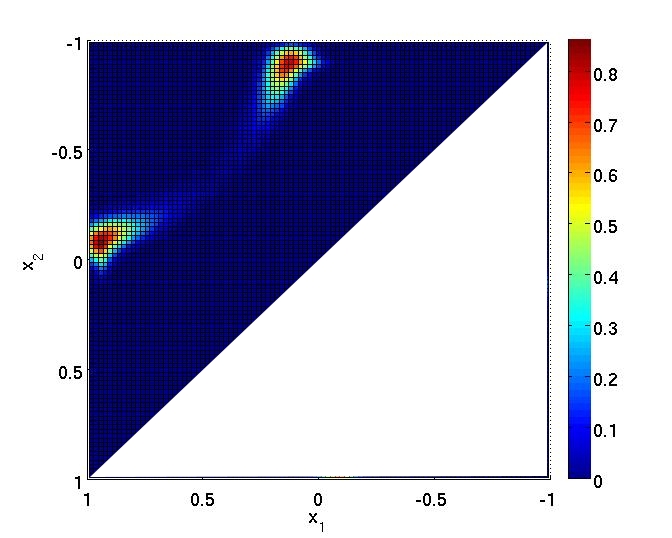}
}
\subfigure[ $\widetilde{P}_2(x_1,x_2)$ ]
{\label{fig:FProb2}
 \includegraphics[width=2.5in]{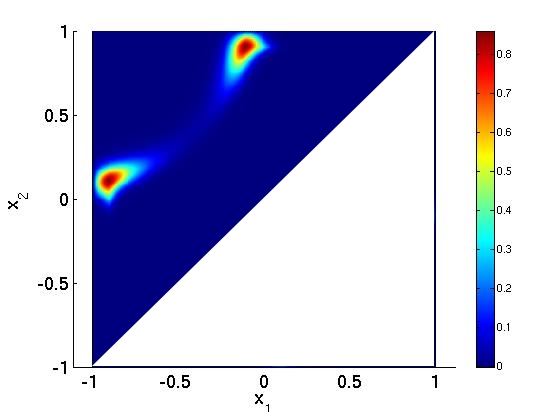}
}
\caption{Using the parameters in \eqref{parameter}, the
  stationary probability density of a $2$-ion channel is computed as 
  {\rm (a)}
  histogram from Brownian dynamics simulation; {\rm (b)} solution of
  \eqref{SSFP1}-\eqref{SSFP5}. Here $x_1$ is the position 
  of the first ion and $x_2$ is the position of the second ion; since we 
  label the ions such that $x_1<x_2$ the state space is a triangle.
}
\label{fig:Prob2}
\end{figure}

\begin{figure}[t]
\centering
\subfigure[ Histogram ]
{\label{fig:SSProb1}
 \includegraphics[width=2.4in, height=2.0in]{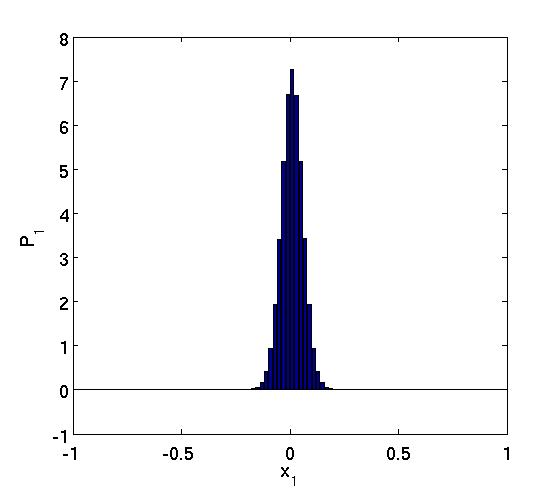}
}
\subfigure[ $\widetilde{P}_1(x_1)$ ]
{\label{fig:FProb1}
 \includegraphics[width=2.4in, height=2.0in]{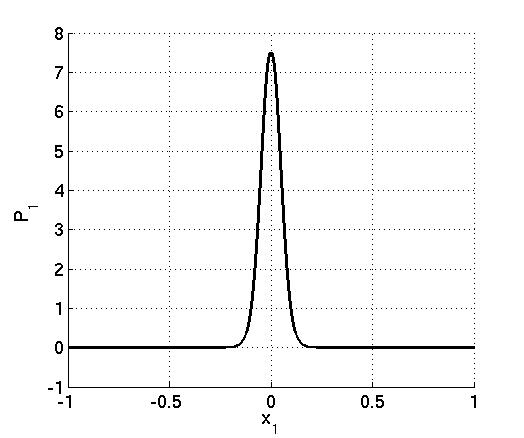}
}
\caption{Using the parameters in \eqref{parameter}, the stationary
  probability density of the one-ion state is computed by {\rm (a)} 
  histogram from
  Brownian dynamics simulation; {\rm (b)} solution of 
  \eqref{SSFP1}-\eqref{SSFP5}. Here $x_1$ is the position 
  of the single ion in the channel.}
\label{fig:Prob1}
\end{figure}

In general the entry rates may be functions of the current ion numbers 
and locations as well as the intracellular $\mathcal{I}$ and 
extracellular $\mathcal{E}$ environments. However, for this illustrative 
example we suppose that they are constant subject to the constraint 
set out above.
Thus we choose
\[ H_0 = \lambda, \quad G_0 = \mu, \quad 
H_1 = \lambda \heaviside(x_1-x_-), \quad
G_1 = \mu \heaviside(x_+-x_1),\]
where $ \heaviside$ is the Heaviside function.
Recall that $H_2 = G_2 = 0$ since the
channel is then fully occupied.
To run the Brownian simulation, we set the time step $\Delta t = 100
\,\mbox{ns}$, and the physical parameters as
\begin{gather}
L = 1 \mbox{nm}, \;\;\;   x_{\pm} = \pm 0.9 \,\mbox{nm}, \;\;\; \xi = 0  
\,\mbox{nm}, \;\;\; d =  0.5\,\mbox{nm},\;\;\;  
D = 1 \mbox{nm}^2\cdot\mbox{ns}^{-1},\quad
\lambda = \mu = 5 \mbox{ns}^{-1}, \nonumber   \\ 
\label{parameter}
T=298\, \mbox{K},\quad 
k_B = 1.38 \times 10^{-23}\, 
\mbox{J} \cdot\mbox{K}^{-1}, \quad 
U = 0 \,\mbox{V}\cdot\mbox{nm}^{-1}, \quad 
Z =  e .
\end{gather}
We use the nanometer as the unit of length and the nanosecond as the
unit of time.
We evolve \eqref{Langevin} for  $2 \times 10^{9}$ iterations  until a
dynamic equilibrium  is reached. During the simulation the number of
ions in the channel varies in time as ions enter and
leave. We record the number of ions and their locations at each time
step. 
We find that the proportion of time spent with $k$ ions in the
channel, $J_k$ say, is given by
 \begin{equation}
\label{BDJ}
 J_0 \approx 0.0000, \quad J_1 \approx 0.8986,\quad J_2 \approx 0.1014.
 \end{equation}
Thus, for these parameters, the channel is almost never empty,
 and for nearly 90\% of the time there is just one ion in the
channel, with two ions the remaining 10\% of the time.
The histograms of $2$-ion distribution and $1$-ion distribution 
are plotted in Fig. \ref{fig:SSProb2} and Fig. \ref{fig:SSProb1},
respectively.

The stationary probability distributions  $\widetilde{P}_2(x_1, x_2)$,
$\widetilde{P}_1(x_1)$ and  $\widetilde{P}_0$ satisfy the stationary
Fokker-Planck equations for a two-ion channel, namely
\begin{subequations}
\label{FP2}
\begin{eqnarray}
0 &=& D \,\nabla
\cdot \left( \nabla \widetilde{P}_2({x}_1, {x}_2) 
 +  \widetilde{P}_2({x}_1, {x}_2) \, \nabla \Phi_2({x}_1,  {x}_2)
 \right)
\nonumber \\  &&\mbox{ }
+  \lambda\heaviside(x_2-x_-)\,\delta({x}_1-x_-)\widetilde{P}_{1}({x}_2) 
 +  \mu\heaviside(x_+-x_1)\, \delta({x}_2-x_+)
 \widetilde{P}_{1}({x}_1), \qquad
\label{SSFP1}\\
0 &=& D \,\fdd{}{x_1}\left( \fdd{\widetilde{P}_1}{x_1}({x}_1) 
 +  \widetilde{P}_1({x}_1) \fdd{\Phi_1}{x_1}({x}_1)  \right)
- \left(\lambda\heaviside(x_1-x_-)+ \mu\heaviside(x_+-x_1)
\right)  \widetilde{P}_1({x}_1) \nonumber \\  
&&\mbox{}+  \lambda \,\delta({x}_1-x_-)\widetilde{P}_{0}  +  \mu\,
\delta({x}_1-x_+) \widetilde{P}_{0}\nonumber\\   
&& 
\mbox{}+ D \left(
\frac{\partial \widetilde{P}_{2}}{\partial x_1}  + 
\widetilde{P}_{2}  \pd{\Phi_2}{x_1}\right)(-L,x_1) - D \left(
\frac{\partial \widetilde{P}_{2}}{\partial x_2}  + 
\widetilde{P}_{2}  \pd{\Phi_2}{x_2}\right)(x_1,L) 
 , \label{SSFP2} \\
0 & = & -(\lambda+\mu)\widetilde{P}_0+ D \left(
\fdd{\widetilde{P}_{1}}{x_1}  + 
\widetilde{P}_{1}  \fdd{\Phi_1}{x_1}\right)(-L) - D \left(
\fdd{\widetilde{P}_{1}}{x_1}  + 
\widetilde{P}_{1}  \fdd{\Phi_1}{x_1}\right)(L).
\end{eqnarray}
with the boundary conditions 
\beq
\widetilde{P}_2(-L,x_2)=\widetilde{P}_2(x_1,L)=0, 
\qquad \widetilde{P}_1(-L)=\widetilde{P}_1(L)=0.
\eeq
and 
\beq
\lim_{x_1 \rightarrow x_2} \left(\pd{\widetilde{P}_2}{x_1} + \widetilde{P}_2
  \pd{\Phi_2}{x_1}\right)  
= \lim_{x_1 \rightarrow x_2} \left(\pd{\widetilde{P}_2}{x_2} + \widetilde{P}_2
  \pd{\Phi_2}{x_2} \right).\label{SSFP5}
\eeq
\end{subequations}

We solve \eqref{SSFP1}-\eqref{SSFP5} by the finite element PDE solver
{\em Comsol} with $28800$ elements. The stationary
distribution $\widetilde{P}_2(x_1,x_2)$ is shown in
Fig. \ref{fig:FProb2}, and $\widetilde{P}_1(x_1)$ is shown in
Fig. \ref{fig:FProb1}. We see that these agree with the histograms in
Fig. \ref{fig:SSProb2} and Fig. \ref{fig:SSProb1} 
obtained from Brownian dynamics simulations.

We see that $\widetilde{P}_2(x_1,x_2)$ is localised around two discrete
states near $(x_-, \xi)$ and $(\xi, x_+)$, while
$\widetilde{P}_1(x_1)$  is localised  around $x_1=\xi$. The most
likely path between the two states of $\widetilde{P}_2(x_1,x_2)$ can
also be faintly seen.

This localisation of  $\widetilde{P}_2$ and $\widetilde{P}_1$ motivates
the definition of a small number of discrete states which the system
can adopt, which is the basis for the  discrete transition  rate
theory described 
in the next section. 

\section{Discrete transition rate theory} \label{sec:rate}

We saw in our two-ion example (Figs. \ref{fig:Prob2} and \ref{fig:Prob1})
that $\widetilde{P}_2$ was mainly localised around
two regions in state space, while $\widetilde{P}_1$ was mainly 
localised around one region.
Suppose, in general, that when there are  $k$ ions in the channel the
stationary probability distribution $\widetilde{P}_k({x}_1,
\ldots, {x}_k)$ 
is mainly  localised around $L_k$ small regions. Let us denote these
regions by 
$$
{S}^{(i)}_k \subset \Gamma_k
\quad
\mbox{for} \;
i=1,\ldots, L_k;
$$ 
then $\widetilde{P}_k
$ is very small outside $\cup_{i=1}^{L_k} {S}^{(i)}_k$, so that 
$$
\int_{\Gamma_k\backslash \cup_{i=1}^{L_k} {S}^{(i)}_k}
\widetilde{P}_k ({x}_1, \ldots, {x}_k) \, \dx_1 \ldots \dx_k \approx 0.
$$
The idea of
discrete rate theory is to replace the continuous variable
$\widetilde{P}_k$ with a set of discrete probabilities
corresponding to the states $S_k^{(i)}$, so that
\[
\widetilde{P}^{(i)}_k=\int_{{S}^{(i)}_k} 
\widetilde{P}_k({x}_1, \ldots, {x}_k)\, \dx_1 \cdots \dx_k
\] 
is the (stationary) probability that $(x_1,\ldots,x_k)  \in S_k^{(i)}$.
Note that $\widetilde{P}^{(i)}_k$  is just a number: it is independent
of spatial variables.  In total there are $L_{\Sigma} = \sum_{k=0}^N
L_k$ states in the channel, and 
the sum of the  probabilities of all $L_{\Sigma}$ states is unity
according to \eqref{sumP},  
that is,  
$$
\sum_{k=0}^N \sum_{i=1}^{L_k} \widetilde{P}^{(i)}_k = 1.
$$ 
We now imagine a Markov chain in which the channel undergoes
transitions from one of these discrete states to another, with the
transition probabilities dependent only on the current state (i.e.  
no past history is involved). This Markov chain is illustrated in
Fig. \ref{fig:FP}. 
Such Markov chains for ion channels have been previously considered
for a single ion in a many-well channel \cite{Abad:2009:NRT}. However,
multiple occupancy of the channel leads to a more complicated
transition structure.

Since only one ion can enter or leave at once (so that
$\widetilde{P}_k$ is coupled only to
 $\widetilde{P}_{k-1}$  and
$\widetilde{P}_{k+1}$) we see that $S^{(i)}_k$ may have
transitions  to  and from only the states $S^{(\cdot)}_k$,
 $S^{(\cdot)}_{k-1}$ and $S^{(\cdot)}_{k+1}$.
The general master equation for the time-dependent probability
$P^{(i)}_k(t)$ is of the form
\begin{eqnarray}
\fdd{}{t} P^{(i)}_k(t)&=& \underbrace{\sum_j \alpha^{(i,j)}_{k-1}
  P^{(j)}_{k-1}(t)   + \sum_l \beta^{(i,l)}_{k+1} P^{(l)}_{k+1}(t) +
  \sum_m \gamma^{(i,m)}_k P^{(m)}_k(t)}_{\mbox{influx}}  \nonumber \\
&& \mbox{ }-  
\underbrace{\left(\sum_j \alpha^{(j,i)}_k + \sum_l \beta^{(l,i)}_k +
    \sum_m \gamma^{(m,i)}_k \right) P^{(i)}_k(t) }_{\mbox{Outflux}} \,, 
 \label{mastereq}
\end{eqnarray}
where $\alpha^{(i,j)}_{k}$ is the transition rate from
$S^{(j)}_{k}$ to $S^{(i)}_{k+1}$, $\beta^{(i,j)}_{k}$  is the
transition rate from 
$S^{(j)}_{k}$ to $S^{(i)}_{k-1}$, and  $\gamma^{(i,j)}_k$
is the transition rate from
$S^{(j)}_{k}$ to $S^{(i)}_{k}$. Thus $\alpha$ describes the
influx of a new ion, $\beta$ describes the loss of an ion to the
intracellular or extracellular environment, and $\gamma$ describes a
hopping of the ions within the channel. 
In fact we expect many of these rates to be zero, since, for example,
when we add a new ion to a   channel it must occupy either the
left-most or rightmost potential well. 

The  entry rates for new ions 
$\alpha^{(i,j)}_k$ may be determined from  $H_k$ and $G_k$, which 
for the present purposes we are assuming are given.
The ion escape rates $\beta_k^{(i,j)}$
and hopping rates $\gamma_k^{(i,j)}$ can be computed from the 
notation of mean
escape time and splitting probability, as described below.

To define the mean escape time we set all the influx probabilities to
zero. We then suppose that the channel initially contains  $k$ ions
located  at positions ${x}_1, \ldots, x_k$. We define the 
mean escape time $\tau_k({x}_1, \ldots, {x}_k)$ to be the average
time before the channel undergoes a transition to a $(k-1)$-ion
configuration, that is, the average time for one ion to leave the
channel.
Using the backward-Kolmogorov equation \cite{Redner:2001:GFP} 
it can be shown that $\tau_k$ satisfies
\begin{subequations}
\label{time}
 \begin{gather}
 \Delta \tau_k - \nabla \Phi_k \cdot \nabla \tau_k  = -
 \frac{1}{D}, \quad ({x}_1, \ldots,{x}_k) \in \Gamma_k \\ 
\tau_k = 0 \;\; \mbox{ if }\;\;   x_1=-L  \mbox{ or }x_k=L.
 \end{gather}
\end{subequations}
Then the mean escape time from state ${S}^{(i)}_k$ is given by
\beq
\tau_k[{S}^{(i)}_k] = \frac{\int_{{S}^{(i)}_k}
  \tau_k P_k\,  \dx_1 \cdots \dx_k }{
  \int_{{S}^{(i)}_k}  P_k
  \, \dx_1 \cdots \dx_k }. \label{taueqn}
\eeq
We now determine a similar expression for $\tau_k[{S}^{(i)}_k]$ using
the discrete transition rate model. Equating the two expressions will
then provide information on the rates $\beta_k^{(i,j)}$ and 
$\gamma_k^{(i,j)}$.

To this end suppose that the channel is initially in the state
${S}^{(i)}_k$, so that $P^{(i)}_k(0) = 1$, $P^{(j)}_m(0) = 0$
otherwise. As before the influx rates $\alpha_k$ are set to be 
zero.
The master equation  \eqref{mastereq} for $P^{(\cdot)}_k$ then
decouples from those for  $P^{(\cdot)}_{k-1}$ and $P^{(\cdot)}_{k+1}$,
and we can solve for $P^{(i)}_{k}$. Given this solution we can
determine the mean escape time $\tau_k[{S}^{(i)}_k]$ as
\begin{equation}
\label{timeP}
\tau_k[{S}^{(i)}_k] =
   \frac{ \displaystyle
    \sum_l\sum_j \int_0^{\infty} t\,\beta^{(l,j)}_k
     P^{(j)}_k(t) \,\dt }{ 
     \displaystyle \sum_l  \sum_j \int_0^{\infty} \beta^{(l,j)}_k
     P^{(j)}_k(t) \,\dt }. 
\end{equation}
In calculating the mean escape time we have not distinguished between
the  case that the first ion leaves from
left end into intracellular electrolyte $\mathcal{I}$ and the case where 
the last ion leaves from right end into extracellular electrolyte 
$\mathcal{E}$.
However, it is important that the discrete state model gets the ratio
of these probabilities correct, since this is what causes a net 
ionic flux through the channel. 
Thus the second piece of information we use to determine the rates
$\beta_k^{(i,j)}$ and $\gamma_k^{(i,j)}$ is the splitting probability
$\rho_k({x}_1, \ldots, {x}_k)$. 
This is defined to be the probability of the first ion to exit was 
${x}_1$ from the left-hand side of the channel,  under the condition
that an ion-escaping event from a $k$-ion to a $(k-1)$-ion channel
has occurred, given that the $k$ ions started in positions $(x_1,
\ldots, x_k)$ initially.
The splitting probability function  $\rho_k$  satisfies,
\begin{subequations}
\label{cond}
 \begin{gather}
 \Delta \, \rho_k - \nabla \Phi_k  \cdot \nabla \rho_k = 0  \qquad
 \mbox{ for }
 ({x}_1, \ldots,{x}_k) \in \Gamma_k \label{condprob:a} \\ 
\rho_k = 1 \;\; \mbox{ on }\;\;  x_1=-L,  \qquad  \rho_k = 0 \;\; \mbox{
  on }\;\; x_{k}=L.
 \end{gather}
\end{subequations}
As with $\tau_k$, we can now calculate the splitting probability for
state ${S}^{(i)}_k$ as
\beq
\rho_k[{S}^{(i)}_k] = \frac{\int_{{S}^{(i)}_k}
  \rho_k P_k \,  \dx_1 \cdots \dx_k }{
  \int_{{S}^{(i)}_k}  P_k
  \, \dx_1 \cdots \dx_k }. \label{rhoeqn}
\eeq
To calculate the splitting probability from the Markov chain we need
to separate $\beta_k^{(l,j)}$ into two individual rates representing 
the case that an ion leaves to the right into
the extracellular domain, and the case than an ion moves to the left
into the intracellular domain, that is, we write
\[ \beta_k^{(l,j)} =  \beta_k^{+(l,j)}+ \beta_k^{-(l,j)}.
\]
Then the probability that an ion escapes to the left given that it
escapes is
\begin{equation}
\label{condP}
\rho_k[{S}^{(i)}_k] =  \frac{ \displaystyle \sum_l \sum_j 
\int_0^{\infty}\beta^{-(l,j)}_k P^{(j)}_k(t) \,\dt }%
{ \displaystyle \sum_l\sum_{j} \int_0^{\infty}
  \beta^{-(l,j)}_k  P^{(j)}_k(t) \,\dt  +  \sum_l\sum_{j} \int_0^{\infty}
  \beta^{+(l,j)}_k  P^{(j)}_k(t) \,\dt }. 
\end{equation}
Note that, as in the case of the escape time $\tau_k$, the right-hand
side depends on ${S}^{(i)}_k$ through the initial condition on
$P_k^{(j)}$.

By equating (\ref{taueqn}) with (\ref{timeP}) and (\ref{rhoeqn}) with
(\ref{condP}) we have a number of equations to help determine the
unknown rates $\beta_k^{(l,j)}$ and $\gamma_k^{(l,j)}$. 
Since only the left-most (respectively right-most) ion can escape from
the left-hand side of the channel (respectively right-hand side), many
of the rates $\beta_k^{(l,j)}$ will infact be zero.
If we still do not have enough equations to determine the remaining
$\beta_k^{(l,j)}$ and $\gamma_k^{(l,j)}$, then it will be necessary 
to determine some of
the transition rates between internal states. Since these do not
involve a change in the number of ions in the channel, they may be
determined by standard techniques.
 
Note that to determine the net flow of ions through the channel we
will also have to distingiush between ion entry from the left and from
the right, that is, we should also split
\[ \alpha_k^{(i,j)} =  \alpha_k^{-(i,j)}+ \alpha_k^{+(i,j)}.\]
However, in most  cases (at least) one of these rates will be zero,
since it is not possible to have the same transition between two states
occuring with an ion entering from either side. The one case where
this is possible is the transition between an empty channel and a
one-ion channel, which occurs in our example below.

\subsection{Example of two-ion channel} 
 
Now we revisit the example of two-ion channel in \Sectionword
\ref{sec:example} and illustrate the rate theory using the
parameters in \eqref{parameter}. 

\begin{figure}[t]
\centerline
{
\includegraphics[width=4.2in]{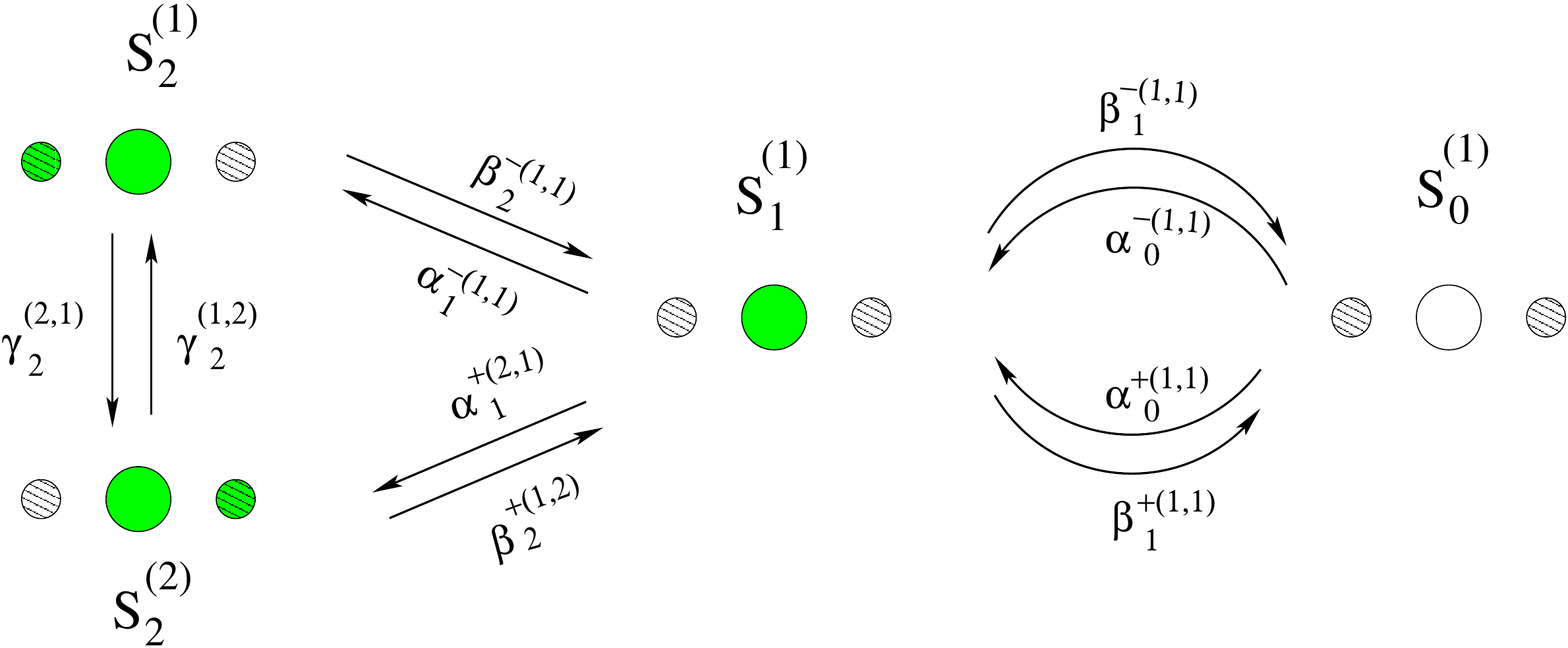}
}
\caption{The transitions between the four different states: the three
  circles represent the left entry point, the binding site and the right
  entry point; a (green) filled circle indicates the presence of an ion.} 
\label{fig:ill1}
\end{figure}

We have seen that the channel can exist in a
$2$-ion, $1$-ion or $0$-ion state.  
From Fig. \ref{fig:Prob1} we see that $\widetilde{P}_1(x_1)$ is 
localised around the
single region $x_1=\xi$, so that there is only one metastable
state with one ion in the channel.
From Fig. \ref{fig:Prob2} we see that $\widetilde{P}_2(x_1,x_2)$ is localised
around the two states $(x_-, \xi)$ and $(\xi, x_+)$. Thus there are
two metastable states with two ions in the channel.
Thus our Markov chain comprises the four states
\begin{equation}
{S}^{(1)}_2: \{(x_-, \xi)\}, \quad {S}^{(2)}_2:
\{(\xi, x_+)\}, \quad  {S}^{(1)}_1: \{\xi\},\quad
{S}^{(1)}_0: \{\}. 
\label{fourstate}
\end{equation}  
Thus $L_2 =
2$, $L_1 = 1$, $L_0 = 1$ and overall there are $L_{\Sigma}=4$ states for
this channel. 
These states, and the transitions between them, are illustrated in
Fig. \ref{fig:ill1}. The circle at center represents the binding
site $x = \xi$, and two other circles 
represent the left and right entry positions $x = x_{\pm}$. A 
(green) filled circle
represents a position occupied by an ion. Note that for the
transitions between $S_0^{(1)}$ and $S_1^{(1)}$ it is important to
distinguish between ions entering and leaving from the right and from
the left, so that we can calculate the net flow of ions through the
channel.

Let us first consider the ion entry rates.
We find
\[
\alpha_1^{+(1,1)}  =  0,\quad
\alpha_1^{-(1,1)}  =  \lambda,\quad
 \alpha_1^{+(2,1)} =  \mu, \quad \alpha_1^{-(2,1)} =  0, \quad
\alpha_0^{+(1,1)}  =   \mu,\quad
\alpha_0^{-(1,1)}  =   \lambda.
\]
Note that the two zero values arise because the transition from
$S_1^{(1)}$ to $S_2^{(1)}$ occurs via 
an ion entering from the left, while that  from $S_1^{(1)}$ to
$S_2^{(2)}$ occurs via 
an ion entering from the right. Note also that $\alpha_2^{(i,j)} = 0$ for
all $i,j$ since with two ions in the channel is already full to capacity.

Let us now consider the ion leaving rates $\beta_k^{(i,j)}$. In principle we
have six of these to determine. However, since the transition from 
 $S_2^{(1)}$ to $S_1^{(1)}$ must occur via an ion leaving from the
 left we know that $\beta_2^{+(1,1)}=0$. Similarly the transition from 
 $S_2^{(2)}$ to $S_1^{(1)}$ must occur via an ion leaving from the
 right, so  we know that $\beta_2^{-(1,2)}=0$.
This leaves $\beta_2^{-(1,1)}$, $\beta_2^{+(1,2)}$, $\beta_1^{-(1,1)}$
and $\beta_1^{+(1,1)}$ to determine. To these we must add the two
hopping rates $\gamma_2^{(1,2)}$ and $\gamma_2^{(2,1)}$.

Denoting the state of the system by the probability vector
$\mathbf{P} = (P^{(1)}_2, P^{(2)}_2, P^{(1)}_1,P^{(1)}_0)^T$, 
the master equation governing the evolution of the Markov chain is
then 
\begin{subequations}
\begin{equation}
\label{transition}
\fdd{\mathbf{P}}{t} = \mathcal{T} \cdot \mathbf{P}(t),
\end{equation}
where the $4 \times 4$ transition matrix $ \mathcal{T}$ is given by
{\small
\begin{equation}
\label{Tmatrix}
\mathcal{T} = \left( \begin{array}{cccc}  -\beta^{-(1,1)}_2 -
    \gamma^{(2,1)}_2 &\gamma^{(1,2)}_2  &\alpha^{-(1,1)}_1 &0 \\[5mm]
    \gamma^{(2,1)}_2  & -\beta^{+(1,2)}_2- \gamma^{(1,2)}_2
    &\alpha^{+(2,1)}_1 &0 \\[5mm]  \beta^{-(1,1)}_2   &\beta^{+(1,2)}_2
    &\parbox{3cm}{$-\alpha^{-(1,1)}_1 - \alpha^{+(2,1)}_1$\\[-1mm]$\mbox{ }-
      \beta^{+(1,1)}_1 - \beta^{-(1,1)}_1$} & \alpha^{+(1,1)}_0 + \alpha^{-(1,1)}_0 \\[5mm] 0 &0 &
    \beta^{+(1,1)}_1+\beta^{-(1,1)}_1 & -\alpha^{+(1,1)}_0 - \alpha^{-(1,1)}_0  
\end{array} \right) .
\end{equation}%
}%
\end{subequations}%
As expected, the sum of each column of the matrix $\mathcal{T}$ is zero
(since the system (\ref{transition}) conserves probability), so the
matrix is rank 
deficient. The stationary probability $\widetilde{\mathbf{P}}$ is the 
eigenvector associated with zero eigenvalue of matrix $\mathcal{T}$.  

\begin{figure}[t]
\centering
\subfigure[$\tau_2(x_1, x_2)$]
{\label{fig:tau2}
 \includegraphics[width=2.45in]{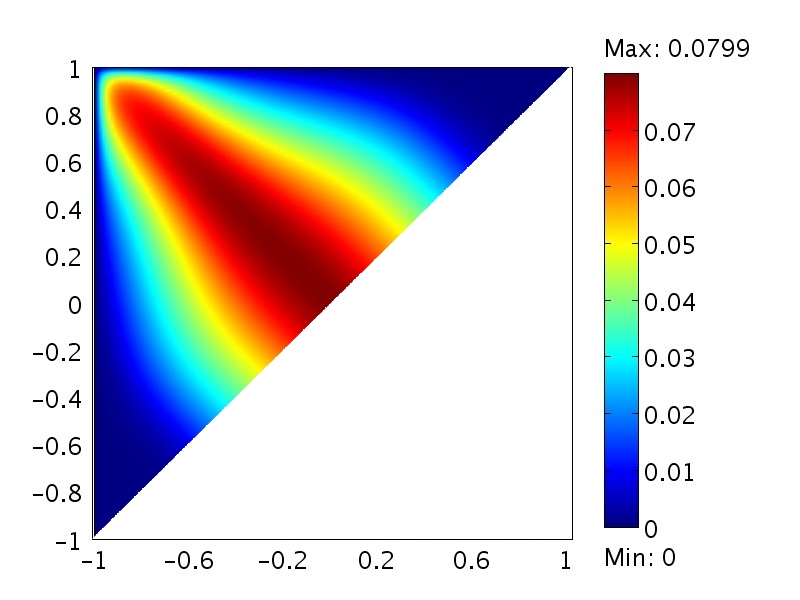}
}
\subfigure[$\rho_2(x_1,x_2)$]
{ \label{fig:P2L}
 \includegraphics[width=2.45in]{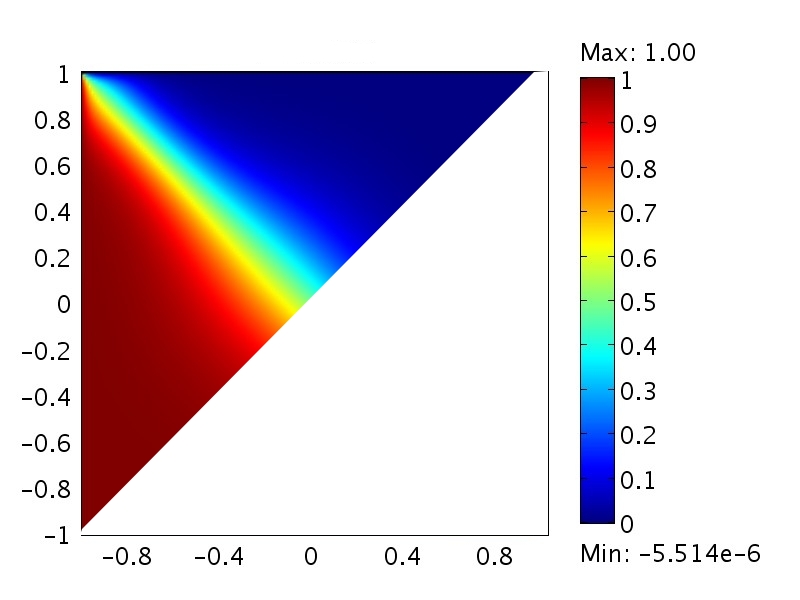}
}
\caption{The parameters are given in \eqref{parameter}. 
{\rm (a)} Mean escape time $\tau_2(x_1, x_2)$ for $2$-ion transiting 
into $1$-ion state. 
{\rm (b)} Splitting probability $\rho_2(x_1, x_2)$ of ion exiting 
from left side under the condition that an ion escaping event occurs.}
\end{figure}

To calculate the mean escape time and splitting probability we set the 
all entry rates to zero and solve (\ref{transition}).  
To emphasize that this is an auxilliary problem and not the true
Markov chain we denote the
probability of lying in each state by $q^{(1)}_2(t), 
q^{(2)}_2(t)$, $q^{(1)}_1(t), q^{(1)}_0(t)$ respectively. Then
(\ref{transition}) is 
\begin{subequations}
\beqa
\fdd{{q}^{(1)}_2}{t}
& =& -\left( \gamma^{(2,1)}_2 + \beta^{-(1,1)}_2 \right)q^{(1)}_2
+ \gamma^{(1,2)}_2 q^{(2)}_2,\label{1}\\ 
\fdd{{q}^{(2)}_2}{t} &=& -\left( \beta^{+(1,2)}_2
+ \gamma^{(1,2)}_2 \right) q^{(2)}_2 + \gamma^{(2,1)}_2 q^{(1)}_2, \label{2}\\ 
\fdd{{q}^{(1)}_1}{t} &=& \beta^{-(1,1)}_2 q^{(1)}_2 + \beta^{+(1,2)}_2
q^{(2)}_2 -\left(\beta^{+(1,1)}_1  + \beta^{-(1,1)}_1
 \right)q^{(1)}_1 , \label{3}
\\
\fdd{{q}^{(1)}_0}{t} &=& \left(\beta^{+(1,1)}_1  + \beta^{-(1,1)}_1
 \right)q^{(1)}_1. \label{4}
\eeqa
\end{subequations}
The first two equations decouple. We start by considering the state 
$S_2^{(1)}$, that is, we solve (\ref{1})--(\ref{2}) subject to the
initial conditions $q^{(1)}_2(0) = 1$ and $q^{(2)}_2(0)=0$. This gives
\beqa
\left( \begin{array}{c} q^{(1)}_2 \\  q^{(2)}_2 \end{array}
\right) &=&  \frac{1}{\,\lambda_1 - \lambda_2} \left( \begin{array}{c}
    \lambda_1 + \beta^{+(1,2)}_2 + \gamma^{(1,2)}_2 \\
    \gamma^{(2,1)}_2 \end{array} \right)   
\exp(\lambda_1\, t)  \non \\
&& \mbox{ }+   \frac{1}{\,\lambda_2 - \lambda_1}
\left( \begin{array}{c}  \lambda_2 + \beta^{+(1,2)}_2 +
    \gamma^{(1,2)}_2\\ \gamma^{(2,1)}_2 \end{array} \right)
\exp(\lambda_2\, t), 
\eeqa
where  $\lambda_1, \lambda_2$ are two eigenvalues satisfying
\begin{eqnarray*}
\lambda_1 + \lambda_2 &=&
-\left(
\beta^{-(1,1)}_2+\beta^{+(1,2)}_2+\gamma^{(2,1)}_2+\gamma^{(1,2)}_2,
\right)\\
\lambda_1 \lambda_2 &=& \beta^{-(1,1)}_2 \beta^{+(1,2)}_2+
\beta^{-(1,1)}_2 \gamma^{(1,2)}_2 + \beta^{+(1,2)}_2
\gamma^{(2,1)}_2.
\end{eqnarray*}  
Using (\ref{timeP}), the mean escape time $\tau_2[{S}^{(1)}_2]$ is
\begin{subequations}
\label{tau2_rho2_tau1_rho1_formula}
\begin{eqnarray}
\tau_{2} [{S}^{(1)}_2] &=& 
\frac{\displaystyle
  \int_0^{\infty} t \left(\beta^{-(1,1)}_2
  q^{(1)}_2(t) + \beta^{+(1,2)}_2 q^{(2)}_2(t) \right) \,\dt}{\displaystyle 
  \int_0^{\infty}
  \beta^{-(1,1)}_2 q^{(1)}_2(t) + \beta^{+(1,2)}_2 q^{(2)}_2(t) \,\dt} 
  \nonumber
  \\
  &=&
\frac{\beta^{+(1,2)}_2 + \gamma^{(2,1)}_2 + \gamma^{(1,2)}_2}{\beta^{+(1,2)}_2
  \gamma^{(2,1)}_2 + \beta^{-(1,1)}_2 \gamma^{(1,2)}_2 + \beta^{-(1,1)}_2
  \beta^{+(1,2)}_2}\,,  
\end{eqnarray}
and, using (\ref{condP}), the splitting probability $\rho_2[{S}^{(1)}_2]$ is
\begin{eqnarray}
\rho_2[{S}^{(1)}_2]  &=& \frac{\displaystyle\int_0^{\infty} 
\beta^{-(1,1)}_2 q^{(1)}_2(t)
  \,\dt}{\displaystyle\int_0^{\infty}  
  \beta^{-(1,1)}_2 q^{(1)}_2(t) + \beta^{+(1,2)}_2
  q^{(2)}_2(t) \,\dt}  \nonumber \\
&=& \frac{\beta^{-(1,1)}_2 \beta^{+(1,2)}_2 + \beta^{-(1,1)}_2
  \gamma^{(1,2)}_2}
{\beta^{-(1,1)}_2 \beta^{+(1,2)}_2 + \beta^{-(1,1)}_2 \gamma^{(1,2)}_2
  + \beta^{+(1,2)}_2 \gamma^{(2,1)}_2}\,. 
\end{eqnarray}
Similarly, by applying the initial conditions $q^{(1)}_2(0) = 0$ and
$q^{(2)}_2(0)=1$, we find 
\begin{eqnarray}
\tau_2 [{S}^{(2)}_2] &=&  
  \frac{\beta^{-(1,1)}_2
  +   \gamma^{(2,1)}_2 + \gamma^{(1,2)}_2}{\beta^{+(1,2)}_2 \gamma^{(2,1)}_2 +
  \beta^{-(1,1)}_2 \gamma^{(1,2)}_2 + \beta^{-(1,1)}_2 \beta^{+(1,2)}_2}\,, \\ 
1-\rho_2[{S}^{(2)}_2] 
&=& \frac{\beta^{-(1,1)}_2 \beta^{+(1,2)}_2 + \beta^{+(1,2)}_2
  \gamma^{(2,1)}_2} {\beta^{-(1,1)}_2 \beta^{+(1,2)}_2 +
  \beta^{-(1,1)}_2 \gamma^{(1,2)}_2 + \beta^{+(1,2)}_2 \gamma^{(2,1)}_2}. 
\end{eqnarray}
Finally we have to consider the escape time and splitting probability
for state ${S}^{(1)}_1$. With the initial condition $q^{(1)}_2(0) =
q^{(2)}_2(0)=q^{(1)}_0(0)=0$, $q^{(1)}_1(0)=1$, equation (\ref{3})
decouples and is easily solved to give
\beqa
\tau_1 [{S}^{(1)}_1]& =&\frac{\displaystyle \int_0^{\infty} t q^{(1)}_1(t)
  \,\dt}{\displaystyle \int_0^{\infty} q^{(1)}_1 \,\dt} = 
  \frac{1}{\beta^{-(1,1)}_1 +  
  \beta^{+(1,1)}_1},\\
\rho_1[{S}^{(1)}_1] &=&
\frac{\,\beta^{-(1,1)}_1}{\,\beta^{-(1,1)}_1 + \beta^{+(1,1)}_1}. 
\eeqa%
\end{subequations}%
The mean escape times $\tau_2,$ $\tau_1$ and the splitting probability
$\rho_2,$ $\rho_1$ 
can be obtained by solving \eqref{time} and \eqref{cond}, respectively
with $k=2,$ $1$ by {\em Comsol}. Fixing $U=0$, $\xi = 0$,  
the mean escape time $\tau_2$ in triangular domain $\Gamma_2$ is
plotted in Fig. \ref{fig:tau2}, the splitting probability $\rho_2$ in
$\Gamma_2$ is plotted in Fig. \ref{fig:P2L}. Since there is no applied
field across the channel ($U=0$), and the potential well is at the
center,  
the external potential is symmetric with $x=0$, therefore both
functions are symmetric with $x_1 + x_2 = 0$.  We take the values of
functions at the centre of the different states and obtain 
\begin{gather*} 
\tau_2 [{S}^{(1)}_2] = \tau_2 (x_-, \xi) = 0.01113, \quad
\rho_2[{S}^{(1)}_2] = \rho_{2} (x_-, \xi) = 0.9964,\\
\tau_2 [{S}^{(2)}_2] = \tau_2 (\xi, x_+)=0.01113,   \quad
 \rho_2[{S}^{(2)}_2] = \rho_2 (\xi, x_+) =0.0036,\\
\tau_1[{S}^{(1)}_1]=\tau_1(\xi)=3.9385 \times 10^3, \quad 
\rho_1[{S}^{(1)}_1]=\rho_1(\xi)=0.5. 
\end{gather*}
Equating these expressions to those of
\eqref{tau2_rho2_tau1_rho1_formula} we find six 
equations for the six unknown rates. Solving these we find
$$
\beta^{-(1,1)}_2 = \beta^{+(1,2)}_2 = 89.8283, \; \gamma^{(2,1)}_2 =
\gamma^{(1,2)}_2 = 0.3361,\; 
\beta^{-(1,1)}_1 = \beta^{+(1,1)}_1 = 1.2695\times 10^{-4}.  
$$
Note that by using the mean escape time and the splitting probability
we have not had to estimate the internal hopping rates $\gamma_2$ from the
Fokker-Planck equation, but have been able to determine them from the
auxillary problems  we have solved. We will see in
\Sectionword \ref{sec:geometry} that this is 
especially useful when the internal states are not so well defined.

Since there is no external potential gradient in this case ($U=0$) the
rates are symmetric, and there is no net flux through the channel.
However, we can already make some observations.
Firstly, the rates $\beta_1^{\pm(1,1)}$ are tiny compared to the others. Thus the
channel will switch between single and double occupancy, but will
almost never be empty of ions. We will confirm this when we consider
the equilibrium occupancy of the channel in the next section.
Secondly, the exit rates $\beta_2^{-(1,1)}$ 
and $\beta^{+(1,2)}_2$ are about 270 times as large as the
hopping rates $\gamma^{(2,1)}_2$ and $\gamma^{(1,2)}_2$ and. 
This means that, for these values of the
parameters, an incoming ion enters and leaves about 270 times before
it manages to replace the bound ion in the potential well at the
centre of the channel.

\subsection{Stationary probability of each state}

Now that we have determined all the transition rates in
Fig. \ref{fig:ill1}, the stationary probability for the number of ions
in the channel can be calculated explicitly as
\begin{equation}
\label{P2P1P0_formula}
\begin{aligned}
\widetilde{P}^{(1)}_2 + \widetilde{P}^{(2)}_2 &= \frac{\alpha^{-(1,1)}_1
  \tau_2[S^{(1)}_2] + \alpha^{+(2,1)}_1
  \tau_2[S^{(2)}_2]}{1+\frac{1}{\tau_1[S^{(1)}_1]}\frac{1}{\alpha^{-(1,1)}_0
    + \alpha^{+(1,1)}_0} + \alpha^{-(1,1)}_1 \tau_2[S^{(1)}_2] +
  \alpha^{+(2,1)}_1 \tau_2[S^{(2)}_2]} \approx 0.1002, \\  
\widetilde{P}^{(1)}_1&=
\frac{1}{1+\frac{1}{\tau_1[S^{(1)}_1]}\frac{1}{\alpha^{-(1,1)}_0
    + \alpha^{+(1,1)}_0} + \alpha^{-(1,1)}_1 \tau_2[S^{(1)}_2] +
  \alpha^{+(2,1)}_1 \tau_2[S^{(2)}_2]} \approx 0.8998, \\  
\widetilde{P}^{(1)}_0&=
\frac{\frac{1}{\tau_1(S^{(1)}_1)}\frac{1}{\alpha^{-(1,1)}_0 +
    \alpha^{+(1,1)}_0}}{1+\frac{1}{\tau_1(S^{(1)}_1)}\frac{1}{\alpha^{-(1,1)}_0
    + \alpha^{+(1,1)}_0} + \alpha^{-(1,1)}_1 \tau_2(S^{(1)}_2) +
  \alpha^{+(2,1)}_1 \tau_2(S^{(2)}_2)} \approx 2.285 \times
10^{-5}.  
\end{aligned}
\end{equation}
We see that these agree well with the probabilities 
obtained from a Brownian dynamics simulation in (\ref{BDJ}).
The same probabilities may be obtained by integrating the solutions of
Fokker-Planck equations over the configuration space, which gives
\begin{eqnarray}
I_2 &\equiv&   \int_{-L}^L \int_{-L}^{x_2} \widetilde{P}_2(x_1, x_2)\,
\dx_1\, 
\dx_2 \approx 0.1001, \nonumber \\
I_1 &\equiv& \int_{-L}^L
\widetilde{P}_1(x_1)\, \dx_1 
\approx 0.8991, 
\qquad
I_0 \equiv \widetilde{P}_0 \approx 8.3 \times
10^{-4}. 
\label{FPprob}
\end{eqnarray}

\section{Current-voltage curve} \label{sec:IV}

The most important characteristic of an ion channel is its
conductance. In this section, we investigate the channel conductance by
examining the current-voltage curve for various values of the channel
parameters.  
 
The potential difference across a channel is usually around $100 \sim 200$
mV. We therefore vary the potential gradient $U$ in the range $[-0.1,
0.1]$ V nm$^{-1}$, which gives a voltage drop in the range $[-200, 200]$ mV
for a channel of length  $2$ nm.   
Following the framework in \Sectionword \ref{sec:rate}, we  compute 
the transition rates corresponding to each given value of $U$.

By examining the transition network in Fig. \ref{fig:ill1}  we see
that there are two different paths which lead to ions moving from the
intracellular (left-hand side) to the extracellular (right-hand side)
domain, namely
\begin{equation*}
\mbox{PATH 1}:\;\; S^{(1)}_1 \xrightarrow{\alpha^{-(1,1)}_1} S^{(1)}_2 
\xrightarrow{\gamma^{(2,1)}_2} S^{(2)}_2 \xrightarrow{\beta^{+(1,2)}_2}  
S^{(1)}_1, \qquad
\mbox{PATH 2}:\;\; S^{(1)}_1 \xrightarrow{\beta^{-(1,1)}_1}  S^{(1)}_0 
\xrightarrow{\alpha^{-(1,1)}_0} S^{(1)}_1.
\end{equation*}
Both paths start with a channel with one ion bound at the potential
well. In Path 1 another  ion first enters the channel from
the left-hand source to produce a two-ion channel. The two ions then hop to 
the right so the new ion lies in the potential well at the centre of the
channel. The ion released from this well then exits the channel at the
right. Thus in Path 1 we can think of a new ion coming in and
knocking the present ion out the other side.
In Path 2 the ion in the channel first leaves from the right to leave 
an empty channel, and then a new ion enters from the left.

By considering the transition rates we can determine the relative
importance of each of these mechanisms. For the parameters in
(\ref{parameter}) the rate $\beta^{-(1,1)}_1$ is tiny and Path 1
dominates the current. Note that at equilibrium the ion flux entering
from the left is balanced by the  flux which leaves from right for
each path, so that 
$$
\alpha^{-(1,1)}_1 \widetilde{P}^{(1)}_1 =  \beta^{+(1,2)}_2
\widetilde{P}^{(2)}_2, \qquad  
\alpha^{-(1,1)}_0 \widetilde{P}^{(1)}_0 =
\beta^{-(1,1)}_1 \widetilde{P}^{(1)}_1.
$$ 
Similarly ions flow from right to left via the paths,
\begin{equation*}
\mbox{PATH 3}:\;\; S^{(1)}_1 \xrightarrow{\alpha^{+(2,1)}_1} S^{(2)}_2 \xrightarrow{\gamma^{(1,2)}_2} S^{(1)}_2 \xrightarrow{\beta^{-(1,1)}_2}  S^{(1)}_1, \qquad
\mbox{PATH 4}:\;\;S^{(1)}_1 \xrightarrow{\beta^{+(1,1)}_1}  S^{(1)}_0\xrightarrow{\alpha^{+(1,1)}_0} S^{(1)}_1,
\end{equation*}
and in equilibrium
$$
\alpha^{+(2,1)}_1 \widetilde{P}^{(1)}_1 =  \beta^{-(1,1)}_2
\widetilde{P}^{(1)}_2, \qquad 
\alpha^{+(1,1)}_0 \widetilde{P}^{(1)}_0 =
\beta^{+(1,1)}_1 \widetilde{P}^{(1)}_1.
$$ 
Combining the current from each path the net current is given by
 \beqa
 \label{current_1well} 
 I &=& e \Big( \alpha^{-(1,1)}_0 \widetilde{P}^{(1)}_0  +
 \alpha^{-(1,1)}_1 \widetilde{P}^{(1)}_1 - \beta^{+(1,1)}_1
 \widetilde{P}^{(1)}_1 - \beta^{-(1,1)}_2 \widetilde{P}^{(1)}_2 \Big)
 \non \\
&=& e \Big( \beta^{-(1,1)}_1 \widetilde{P}^{(1)}_1 + \beta^{+(1,2)}_2
\widetilde{P}^{(2)}_2 - \alpha^{+(1,1)}_0 \widetilde{P}^{(1)}_0 -
\alpha^{+(2,1)}_1 \widetilde{P}^{(1)}_1\Big) , 
\eeqa
\begin{figure}[t]
\centering
\subfigure[]
{\label{fig:IVplot:lam1}
\includegraphics[width=2.45in]{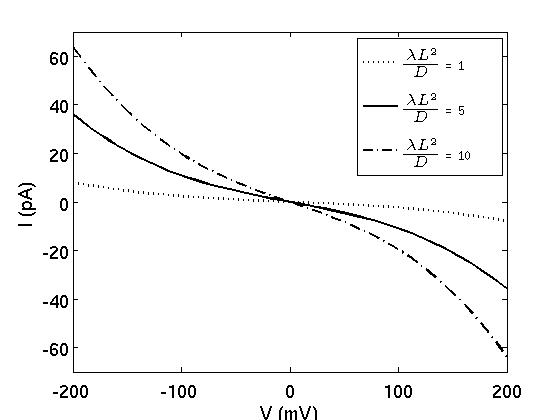}}
\subfigure[]
{\label{fig:IVplot:lam2}
\includegraphics[width=2.45in]{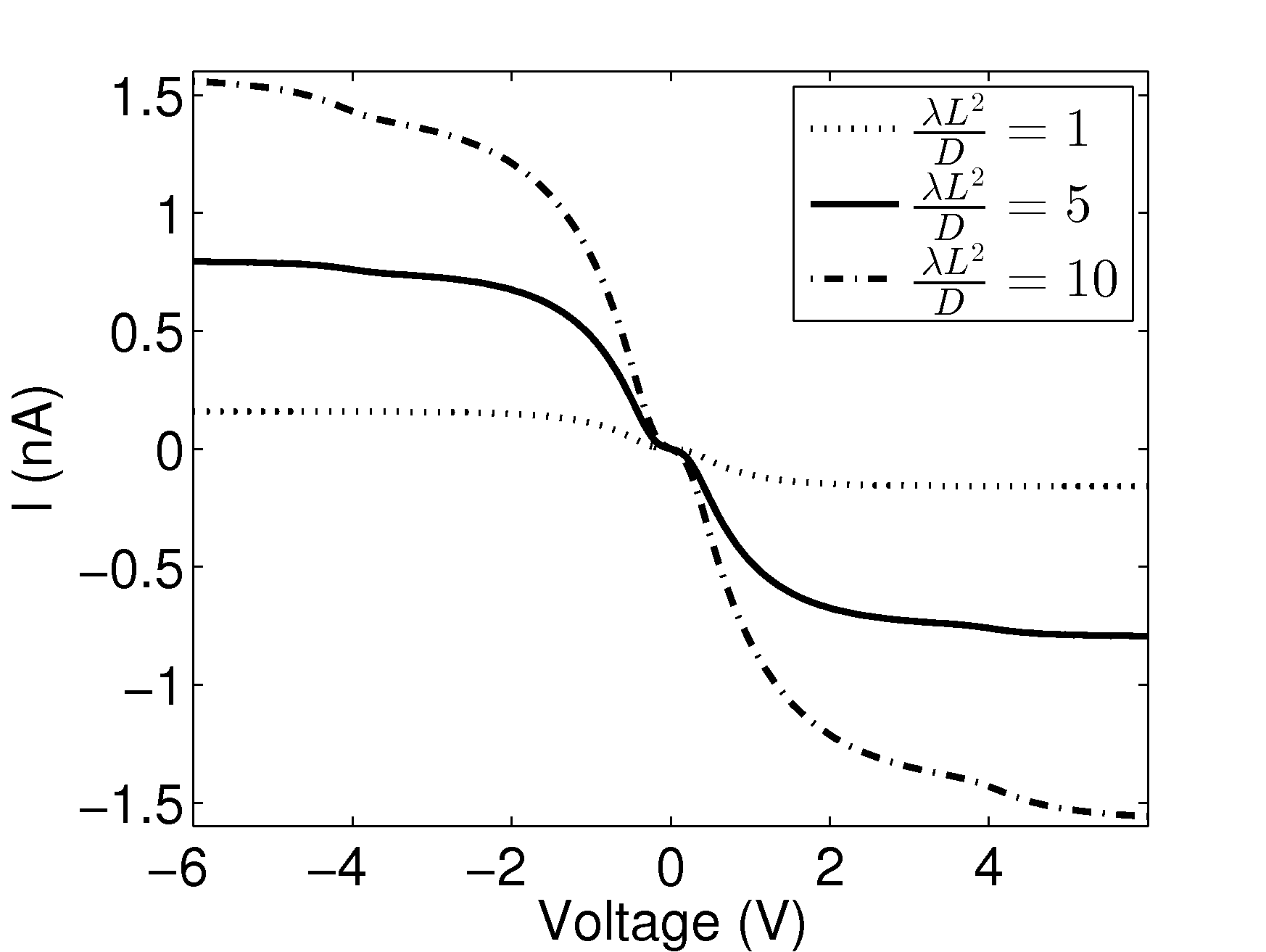}}
\caption{
I-V curve for $L^2\lambda/D = L^2\mu/D=1$, $5$, $10$ with 
$d/L=0.5$; all other
parameters are as  in \eqref{parameter}. {\rm (a)} for small 
voltages, for
which internal transitions in the channel are rate limiting;
{\rm (b)} for larger voltages, at which saturation 
occurs due to the finite rate of entry of ions from the bulk.}
\label{fig:IVplot:lam}
\end{figure}%
\begin{figure}[t]
\centering
{
\includegraphics[width=2.45in]{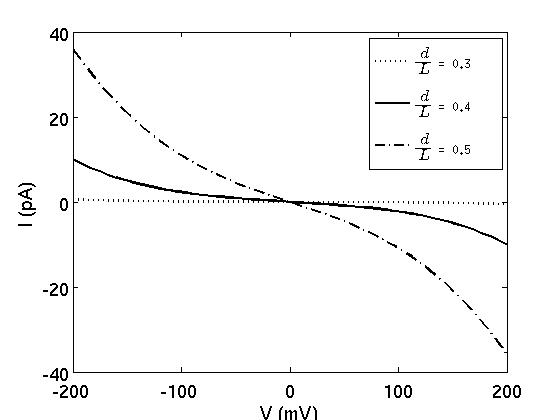}}
\caption{I-V curve for $d/L=0.3$, $0.4$, $0.5$ with $L^2\lambda/D
=L^2\mu/D=5$. All other parameters are as  in \eqref{parameter}.}
\label{fig:IVplot:d}
\end{figure}%
Next we study how the conductance of the channel varies with the
 external potential energy parameter $d/L$
and the dimensionless entry rates $L^2\lambda/D$ and
$\mu L^2/D$.  
We plot  in Fig.~\ref{fig:IVplot:lam} the current-voltage (I-V) curve 
for different entry rates
when $d/L=0.5$. For large voltages the current saturates since it is
limited by the entry rate of ions $\lambda$ and $\mu$; this effect is
illustrated in Fig.~\ref{fig:IVplot:lam2}.
In
Fig.~\ref{fig:IVplot:d} we plot the I-V curve for various values of $d$
with a fixed entry rate.
The slopes of these curves give the conductance of channel, which
grows initially with increasing voltage until diminishing as
saturation sets in. We see that  
as the dimensionless entry rate increases, 
or the potential well gets shallower,
the conductance of the channel increases.     
 
\section{Optimal geometry of potential} \label{sec:geometry}

\begin{figure}[t]
\centering
\subfigure[$\Phi$ vs. $x$]
{ \label{fig:pot_group_1well:a}
\includegraphics[width=2.45in]{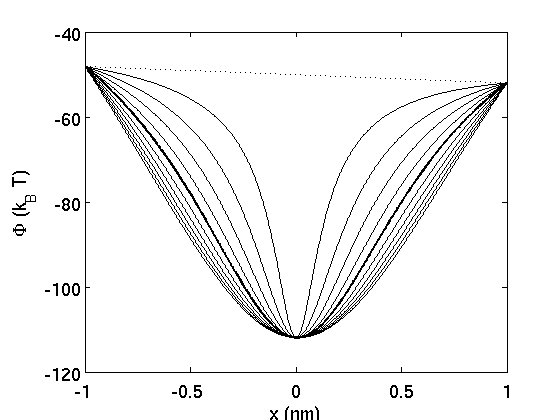}}
\subfigure[$Z$ vs. $d$]
{\label{fig:pot_group_1well:b}
 \includegraphics[width=2.45in]{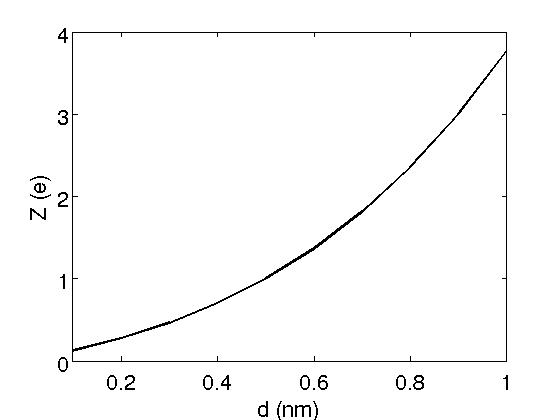}}
\label{fig:pot_gropu_1well}
\caption{{\rm (a)} A group of potential energy wells for $d = 0.1, 0.2, 0.3,
  \ldots,0.9, 1.0$ nm are plotted by solid curves respectively from
  narrow to wide well, the darker curve corresponds to $Z = e$,
  $d/L=0.5$, and the dotted curve corresponds the energy drop due to
  applied field $U=-0.05$  V nm$^{-1}$.  
{\rm (b)} The external charge $Z$ as a function of $d$. }
\end{figure}

We showed that a channel with shallower potential well has larger
conductivity in Fig. \ref{fig:IVplot:d}, if all other physical
parameters are fixed. 
However, the depth of the potential well is not the only factor that
determines the ion flux. Abad et al \cite{Abad:2009:NRT} studied 
the flux through a channel of capacity $N=1$ 
with symmetric M-shape potential energy, and showed that there exists
a critial ratio $\sigma$ of the width of potential well over the
length of channel, at which 
the flux is maximized. This optimal geometry of potential energy
requires the potential well is neither too narrow ($\sigma \to 0$) nor
too wide ($\sigma \to 1$).  

We now perform a similar analysis of the multi-ion channel. 
For our $N=2$ example we study how the shape of an external potential  
on the protein boundary  affects the conductivity of the
channel. We vary the distance $d$ and carefully choose the external
charge as
\[
Z =  e\, \frac{\displaystyle 2-\left(1+0.5^2\right)^{-0.5}}{\displaystyle 
d^{-1}L -\left(1+d^2L^{-2}\right)^{-0.5}},
\] 
 so that the depths 
of the resulting potential wells are all the same (and so that $Z=e$,
when $d/L=0.5$); the variation of the potential well with $d$ is shown in
Fig. \ref{fig:pot_group_1well:b}.  
The resulting group of potential wells with applied field $U=-0.05$ V
nm$^{-1}$  are plotted in Fig. \ref{fig:pot_group_1well:a}, the darker
curve corresponds to $Z=e$, $d=0.5 L$,
the dotted curve shows how the potential wells tilt with the applied field.
Obviously, when $d$ is small, we have a steep potential drop 
near the binding site, and a relatively flat energy landscape near the ends
of the channel. When $d/L$ gets large, the potential well tends to  a
curve 
that is steeper near the end of the channel and flatter near the binding site.

\begin{figure}[htbp]
\centering
\subfigure[$d=0.1$ nm]
{\label{fig:P2:a}
 \includegraphics[width=1.65in]{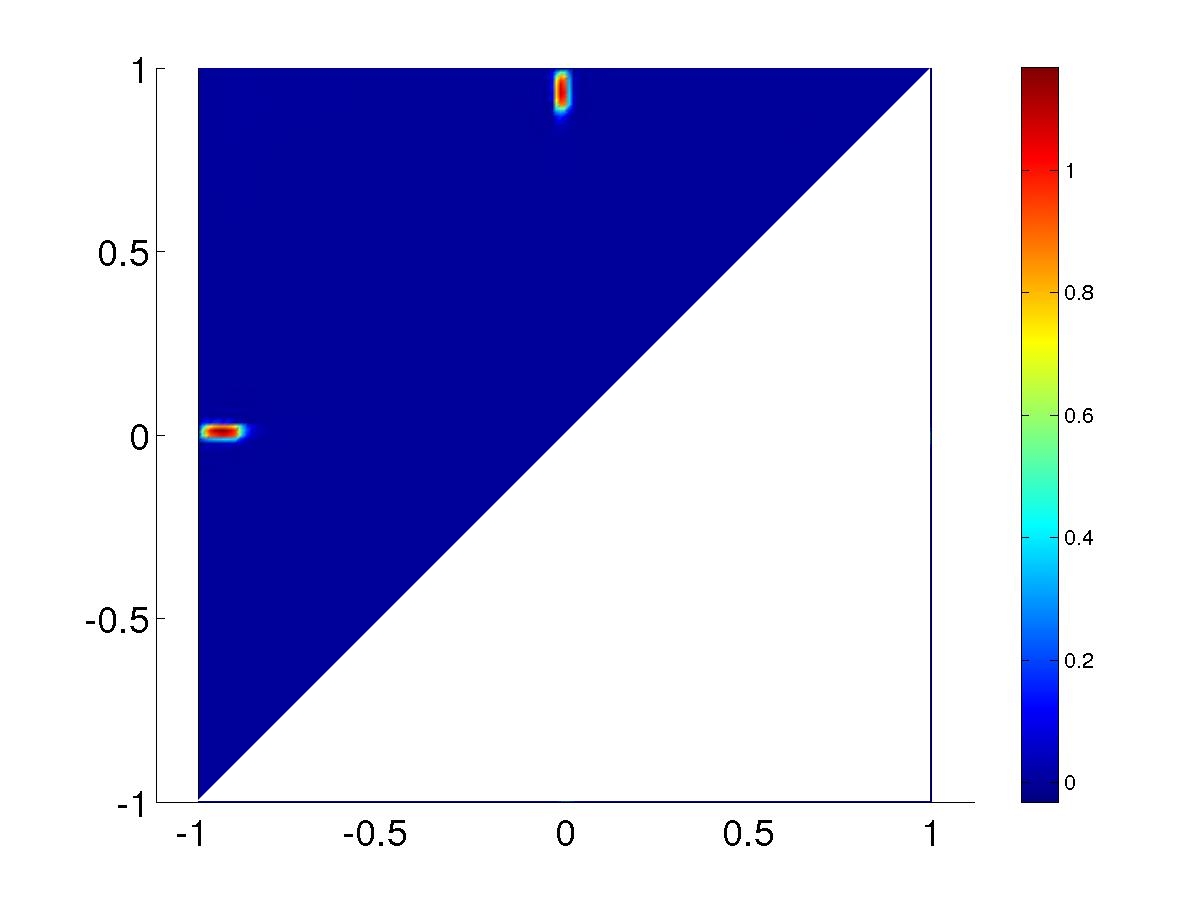}}
\subfigure[$d=0.2$ nm]
{\label{fig:P2:b}
 \includegraphics[width=1.65in]{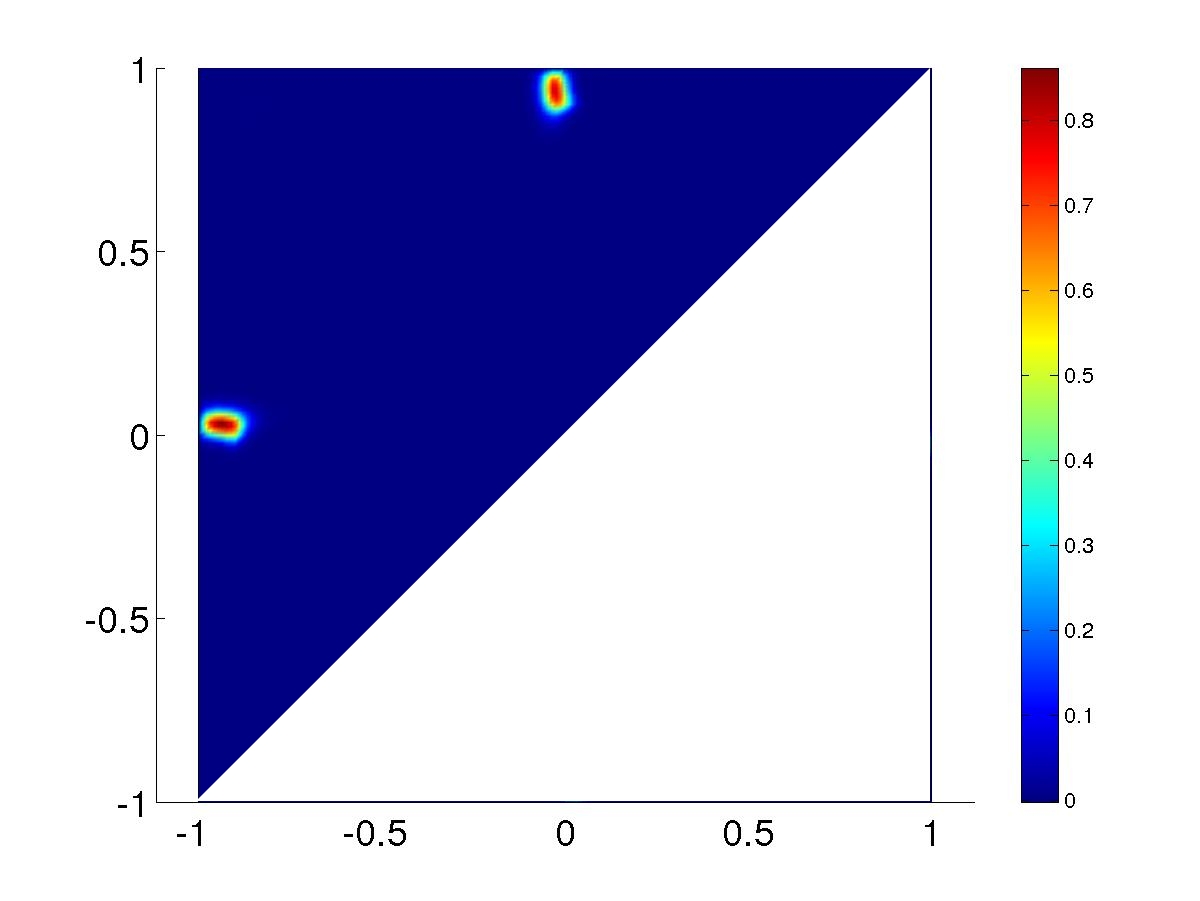}}
\subfigure[$d=0.3$ nm]
{\label{fig:P2:c}
 \includegraphics[width=1.65in]{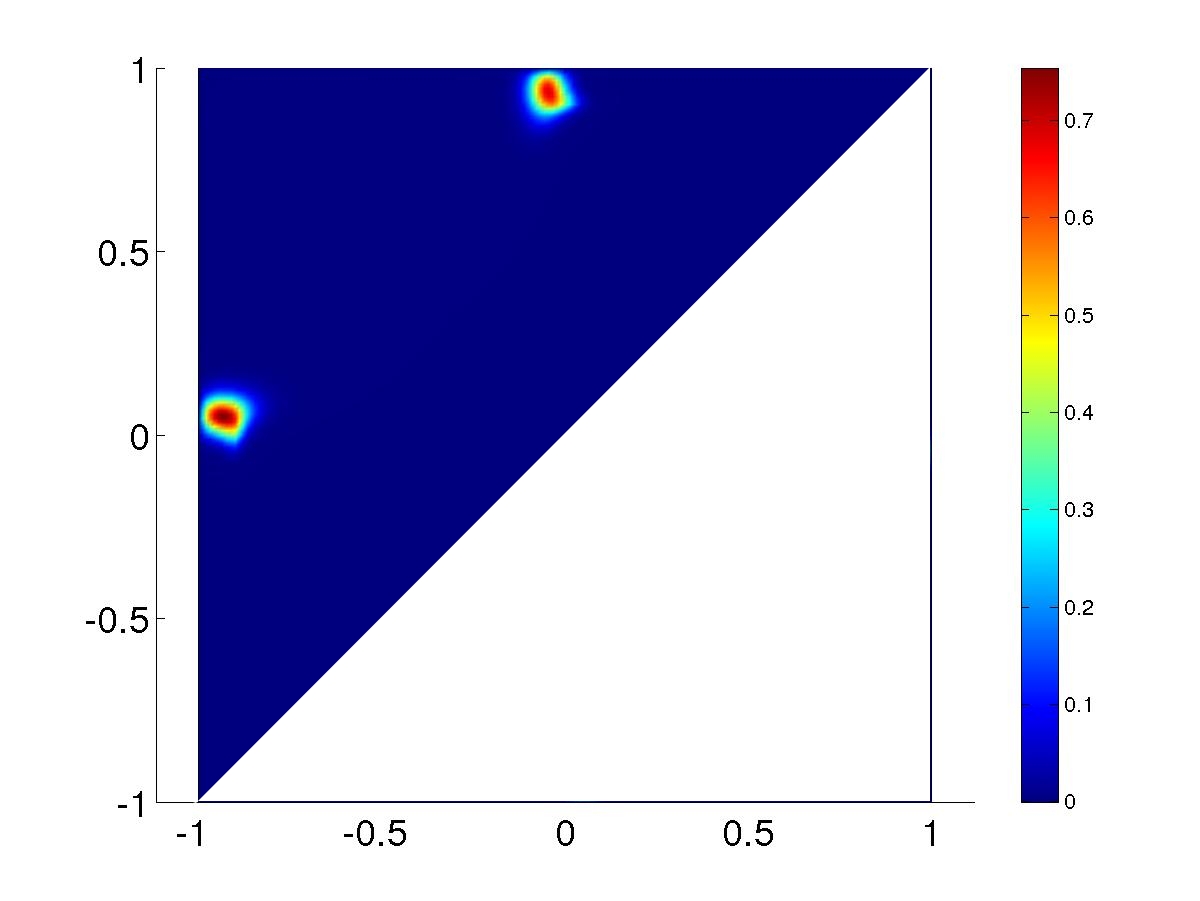}}
\subfigure[$d=0.4$ nm]
{\label{fig:P2:d}
 \includegraphics[width=1.65in]{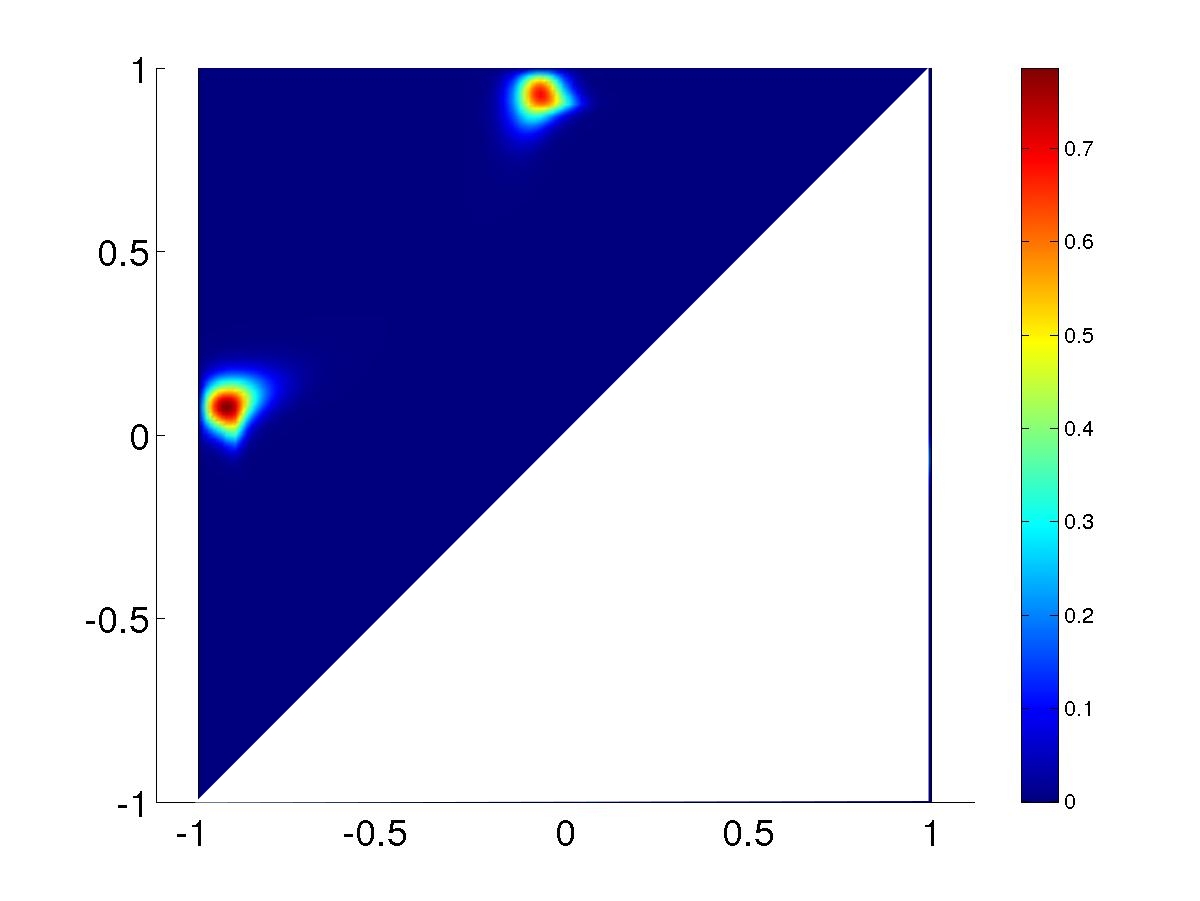}}
\subfigure[$d=0.5$ nm]
{\label{fig:P2:e}
 \includegraphics[width=1.65in]{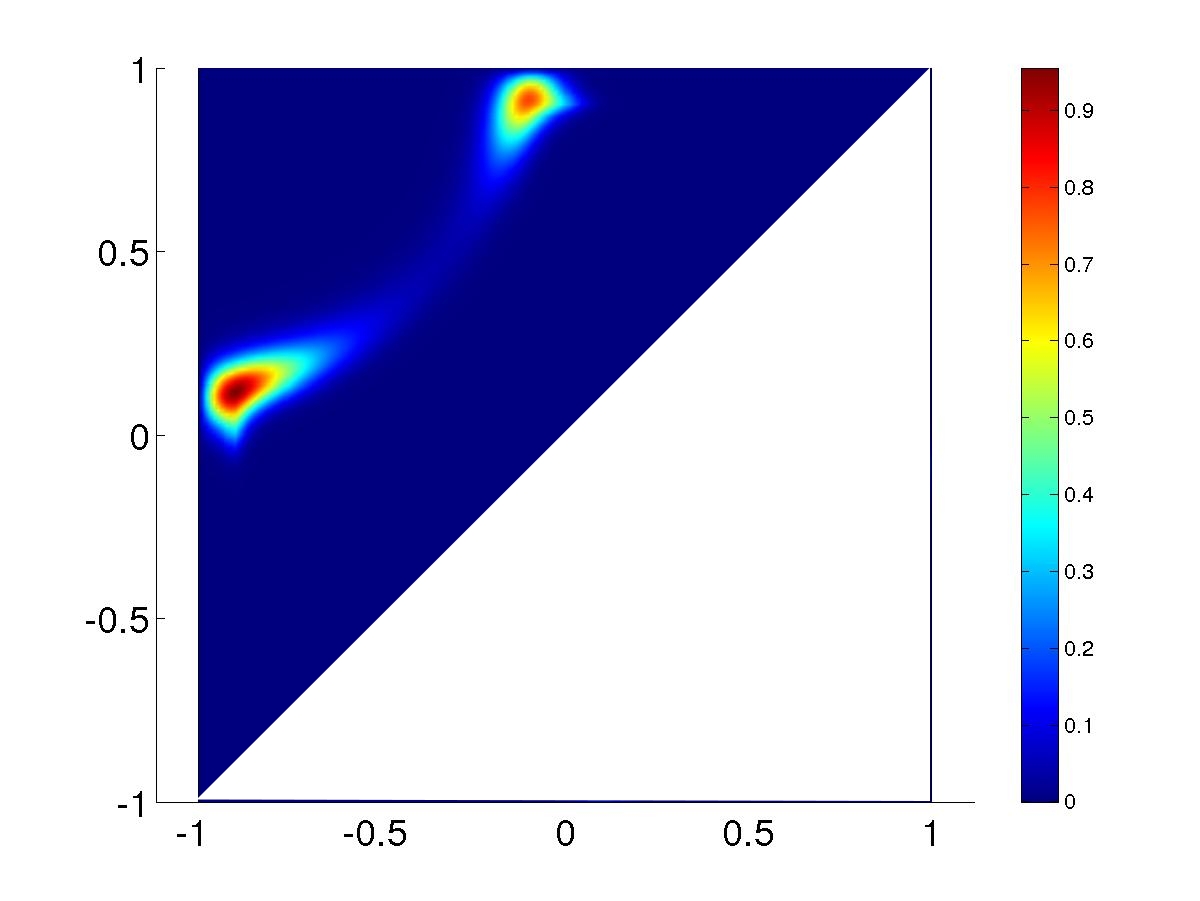}}
\subfigure[$d=0.6$ nm]
{\label{fig:P2:f}
 \includegraphics[width=1.65in]{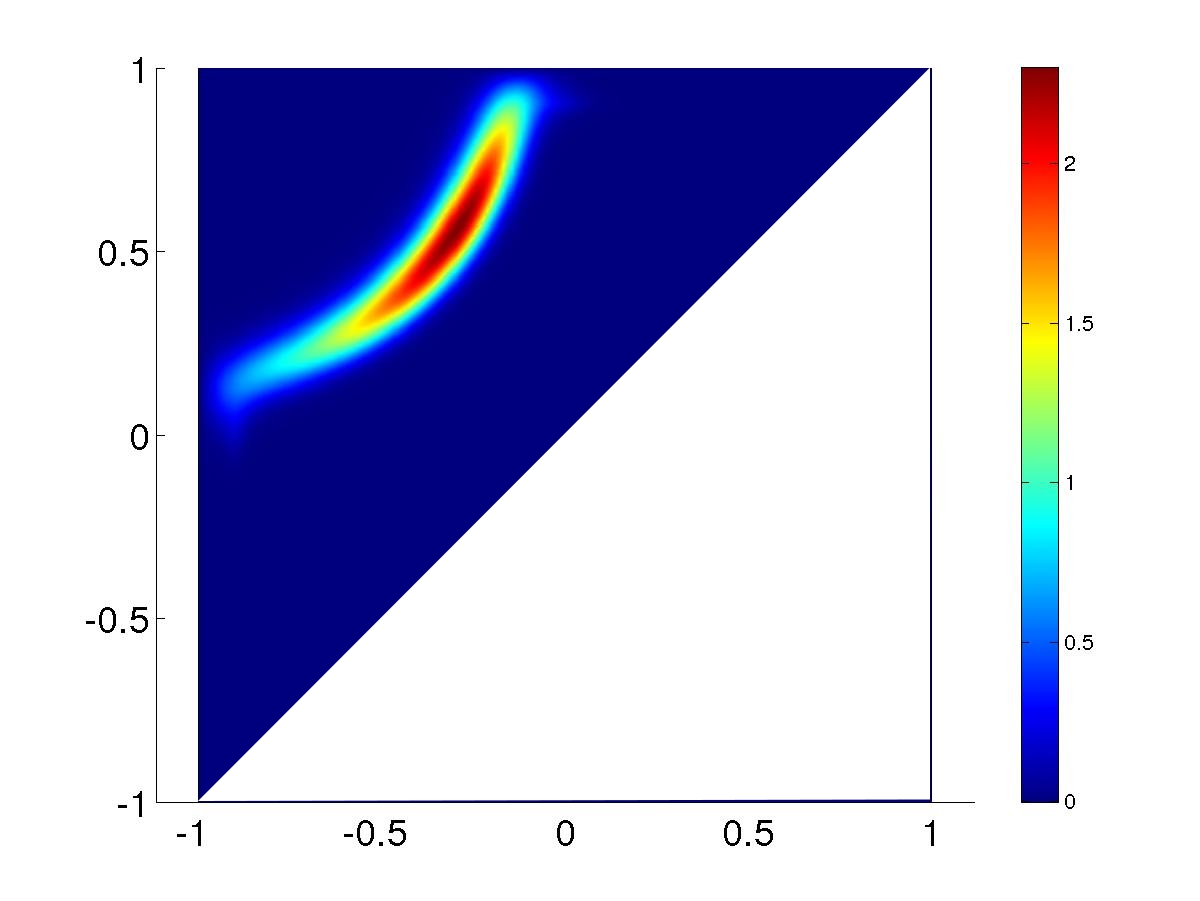}}
\subfigure[$d=0.7$ nm]
{\label{fig:P2:g}
 \includegraphics[width=1.65in]{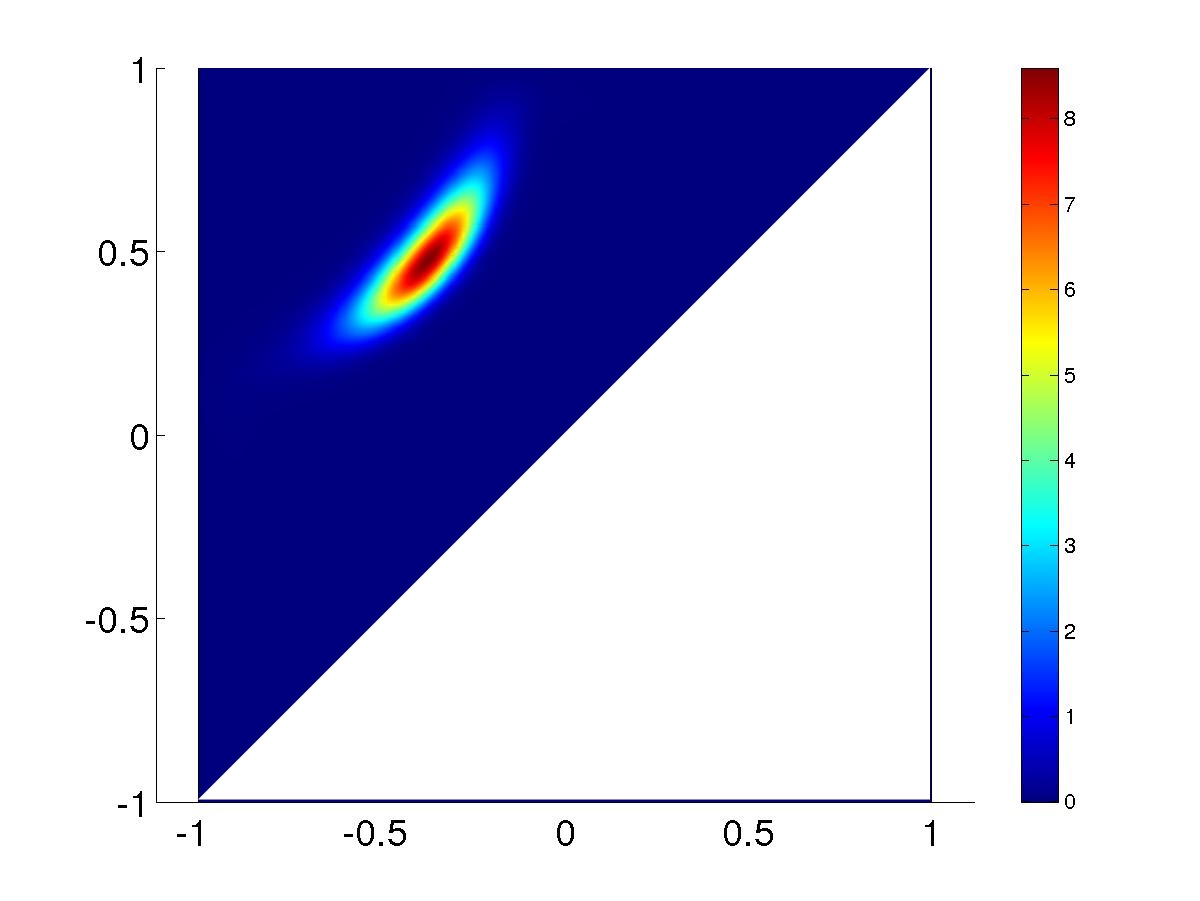}}
\subfigure[$d=0.8$ nm]
{\label{fig:P2:h}
 \includegraphics[width=1.65in]{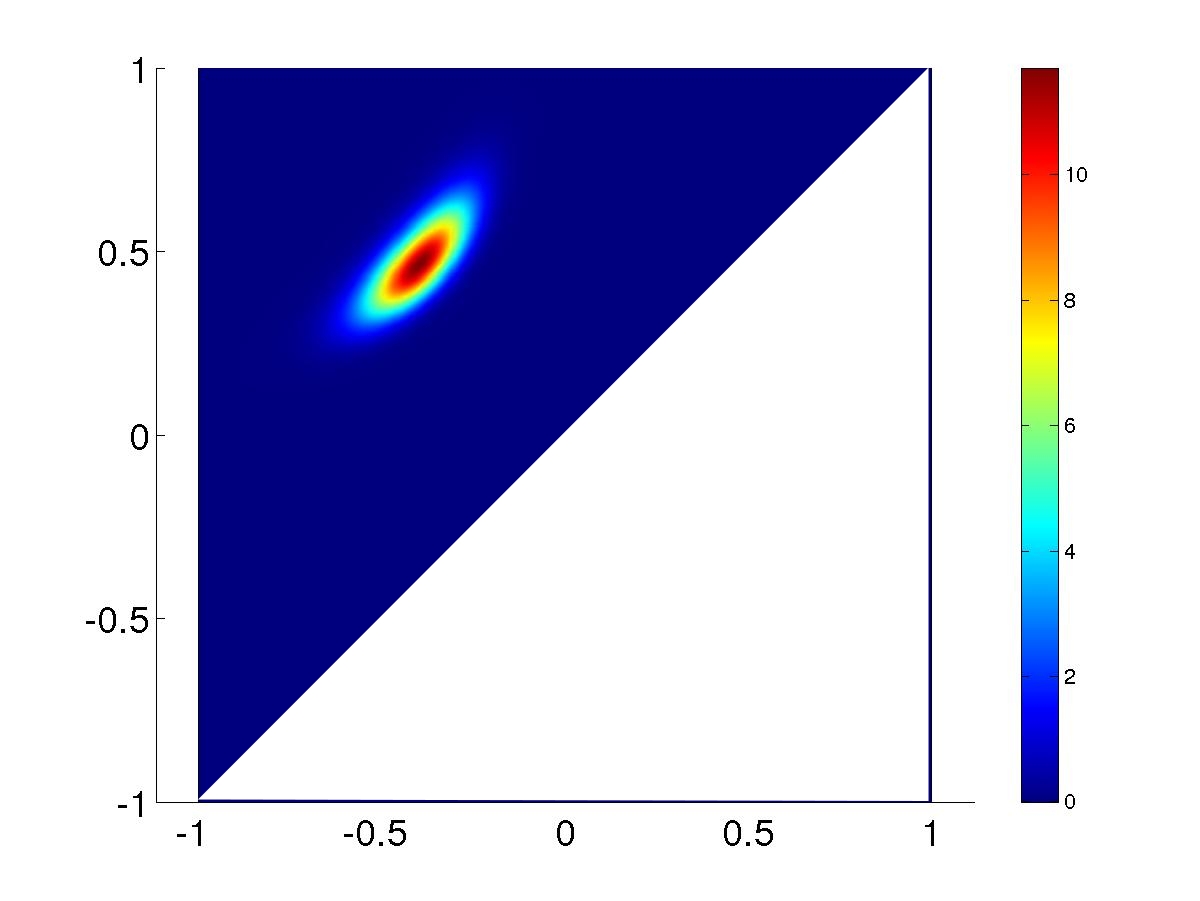}}
\subfigure[$d=0.9$ nm]
{\label{fig:P2:i}
 \includegraphics[width=1.65in]{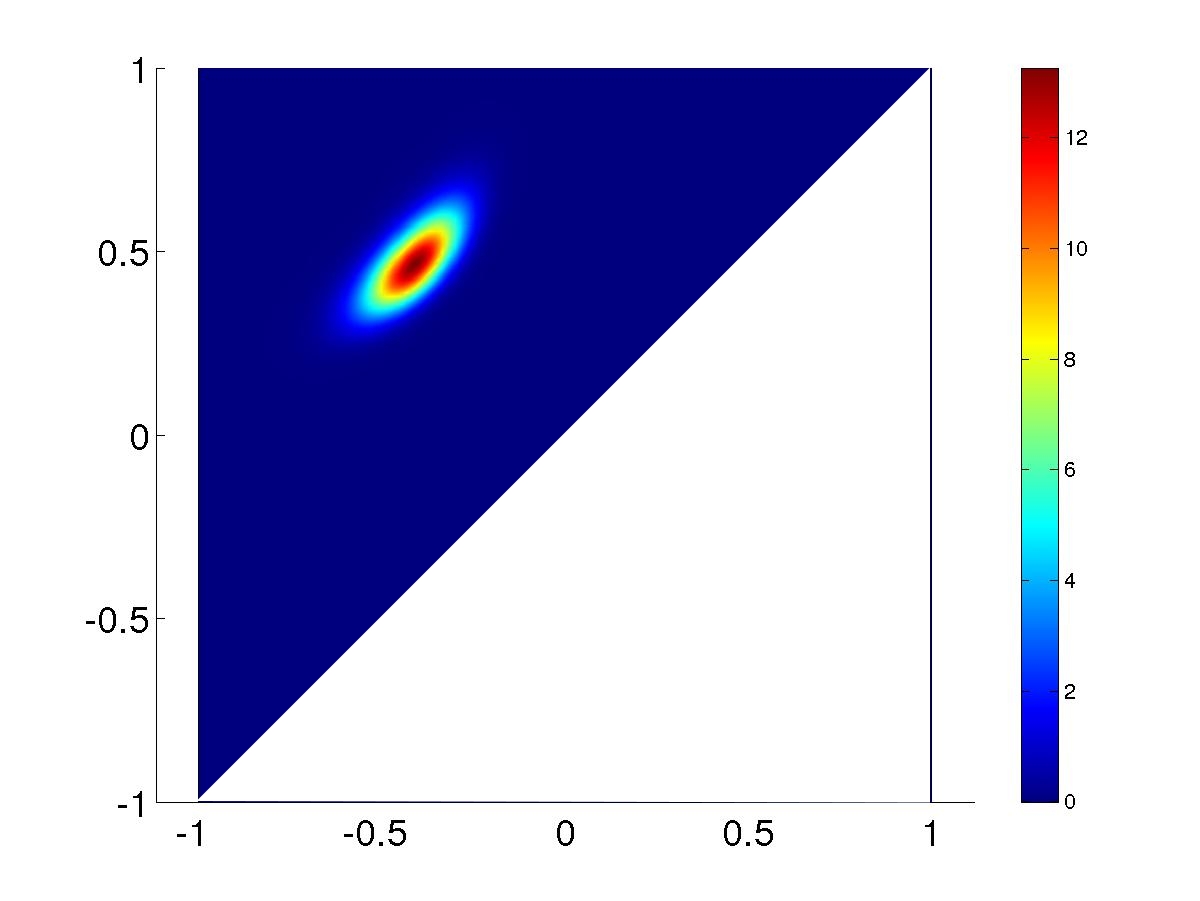}}
\caption{Fixing $\lambda=\mu=5$ ns$^{-1}$, $U=-0.05$ V nm$^{-1}$, we
  plot the stationary probability distribution $\widetilde{P}_2(x,y)$
  for $d =0.1, 0.2, 0.3$,  
$0.4, 0.5, 0.6$, $0.7, 0.8, 0.9$ nm, which are obtained by solving
\eqref{FP2} numerically. 
It shows the change from two distinct states at $(-0.9, 0)$ and $(0, 0.9)$
to a ridge of high probability regime to one distinct state near
$(-0.45, 0.45)$, with the change of geometry of potential well.}
\label{fig:P2}
\end{figure}

Now we fix $U=-0.05$ V nm$^{-1}$, $\lambda= \mu=5$ ns$^{-1}$, and plot
in Fig. \ref{fig:P2} the stationary probability density function
$\widetilde{P}_2(x_1, x_2)$ obtained by solving  
stationary Fokker-Planck equations numerically 
for various values of $d$. Recall that since $x_1
< x_2$, we only look at the upper left triangular domain. 
We observe that when the potential well is very narrow and steep at
the binding site with $d=0.1$ nm, the probability distribution of two
ions  
is localized at two tiny spots around  $(x_-, \xi)$ and $(\xi, x_+)$.  
As $d$ increases until $d=0.5$ nm, the two spots grow a little larger and
shift slightly away from  
$(x_-, \xi)$ and $(\xi, x_+)$. Thus for $d \in [0.1, 0.5]$ nm we can
justify the use of our four-state rate theory. 
For $d$ between $0.5$ nm and $0.6$ nm, a ridge of high probability 
distribution
emerges that connects previous two spots at its two tails.  
This implies that instead of the two ions being trapped  at one of two
states, the two can wander back and forth freely between these two
states.
As $d$ increases even further to $d=0.9$ nm, the ridge shrinks to its
center peak, which corresponds to the two ions both sitting in the
potential well. Thus, for larger values of $d$, we have effectively
only one state for the two-ion occupied channel.

\begin{figure}[t]
\label{fig:S3S4_cmp}
\centering
\subfigure[]
{ \label{fig:tau2_d}
 \includegraphics[width=2.45in]{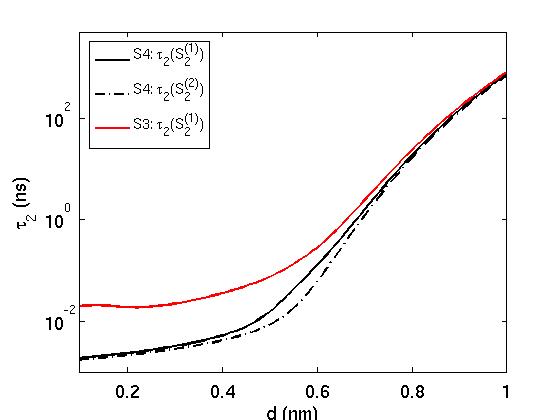}}
\subfigure[]
{ \label{fig:rho2_d}
 \includegraphics[width=2.45in]{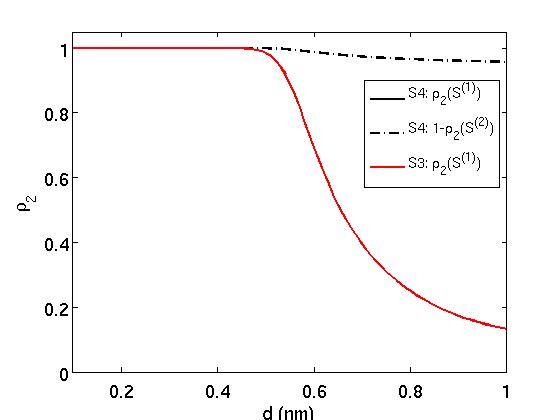}}
\subfigure[]
{ \label{fig:beta_1well}
 \includegraphics[width=2.45in]{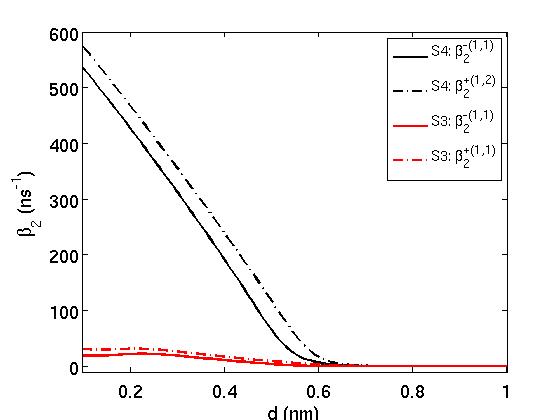}}
\subfigure[]
{ \label{fig:gamma_1well}
 \includegraphics[width=2.45in]{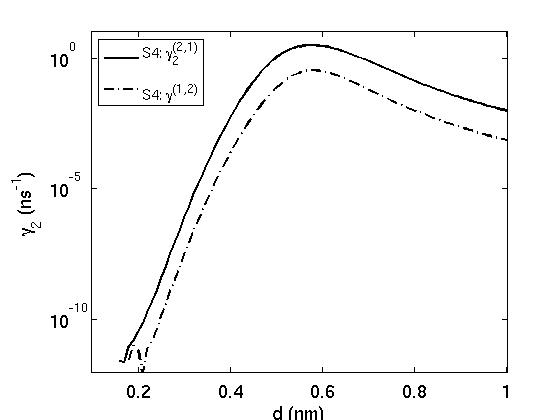}}
\caption{Fixing $U=-0.05$ V nm$^{-1}$, we numerically solve
\eqref{time} and \eqref{condP} and obtain 
the mean escape times and left splitting probabilities at two 
states $S_2^{(1)}=(-0.9,0)$ and $S_2^{(2)}=(0,0.9)$ for 
$d$ between 0.1 nm and 1 nm, which
determine the escaping rates and transition rates by 
\eqref{tau2_rho2_tau1_rho1_formula}. We compare it with the results
of the three-state simplified model $(\ref{threestate})$.
{\rm (a)} The mean escape time $\tau_2$ vs. $d$ at different states.
{\rm (b)} The splitting probability $\rho_2$ vs. $d$ at different states. 
{\rm (c)} The escaping rates $\beta^{(1)}=\beta_2^{-(1,1)}$ 
and $\beta^{(2)}=\beta_2^{+(1,2)}$.
{\rm (d)} The transition rates $\gamma^{(1)}=\gamma_2^{(2,1)}$ 
and $\gamma^{(2)}=\gamma_2^{(1,2)}$.
Note that those rates only depend on the mean escape time and 
left splitting probability, and are independent of entry rates.
}
\end{figure}

Thus, for large $d$, we can define a single-chained three-state system
\begin{equation}
{S}^{(1)}_2: \{((x_-+\xi)/2, (\xi+x_+)/2)\}, \quad
{S}^{(1)}_1: \{\xi \}, \quad 
{S}^{(1)}_0: \{\},
\label{threestate}
\end{equation}
which is similar to four-state system in Fig. \ref{fig:ill1}, 
except that ${S}^{(1)}_2 = {S}^{(2)}_2$ are combined 
and there is no hopping 
$\gamma^{(1,2)}_2$ and $\gamma^{(2,1)}_2$ between them. 
Following the framework in \Sectionword \ref{sec:rate}, all 
the entry rates $\alpha^{-(1,1)}_0 = \alpha^{-(1,1)}_1 = \mu$ 
and $\alpha^{+(1,1)}_0 = \alpha^{+(1,1)}_1 = \lambda$ are 
prescribed, and the escaping rates are easily calculated as
\[ \beta^{-(1,1)}_k = \frac{\,\rho_k(S^{(1)}_k)}{\,\tau_k(S^{(1)}_k)}, 
\quad   \beta^{+(1,1)}_k = \frac{1-\rho_k(S^{(1)}_k)}{\,\tau_k(S^{(1)}_k)} ,    
\quad k=1, 2. 
\]
In Fig. \ref{fig:tau2_d}, we plot the mean escape time $\tau_2$ from
the left state $S_2^{(1)}=(-0.9,0)$ (black solid curve) and the right
state $S_2^{(2)}=(0,0.9)$ (black dash-dotted curve)  
in four-state formulation, and compare them with $\tau_2$ from the
balanced state $S_2^{(1)}=(-0.45, 0.45)$ (red solid curve) in the
three-state formulation (\ref{threestate}). 
Since the potential well is broader as $d$  increases, the second
ion (the one which is not trapped in the well) feels its effect more,
resulting in an exponential increase in the mean escape time. In a
channel with descending 
voltage from left to right ($U=-0.05$ V nm$^{-1}$), it takes a longer
time for two ions in the  left state $(-0.9, 0)$ to escape than two
ions in the 
right state $(0, 0.9)$, so the black solid  
curve is above the black dashed curve. 
For two ions in the balanced state $S_2^{(1)}=(-0.45, 0.45)$ in the
middle of channel, it takes an even longer time, so  the red solid
curve is above the  two black curves. 

Notice that the ratio $\tau_2(-0.9,0)/\tau_2(0, 0.9)$ is
largest when $0.5 < d < 0.6$.   Note also that for small $d$ the
three-state formulation is invalid (so $\tau_2(-0.45, 0.45)$ is
meaningless) but that for large $d$ the ratio $\tau_2(-0.45,
0.45)/\tau_2(-0.9,0)$ is close to $1$: for a broad potential the
mean escape time is insensitive to the precise initial
position of the two ions.

In Fig. \ref{fig:rho2_d} we plot the left splitting probability
$\rho_2$ at the left state $S_2^{(1)}=(-0.9,0)$ (black solid curve)
and the right splitting probability $1-\rho_2$  
at the right state $S_2^{(2)}=(0,0.9)$ (black dash-dotted curve) in
four-state formulation. For large $d$, the external potential has a
large gradient near both  
ends of the channel (as shown in Fig. \ref{fig:pot_group_1well:a}),
which pulls the new ion introduced at either source towards the center
of channel, and into balance with the existing ion  
at around $(-0.45, 0.45)$. So new entering ions at each source are
less likely to leave from the same end of the channel, and both the left
splitting probability at  
left state (black solid curve) and the right splitting probability at
right state (black dash-dotted curve) decrease monotonically as $d$
increases. In addition, the left-splitting  
probability $\rho_2$ at the balanced state $S_2^{(1)}=(-0.45, 0.45)$ 
in the three-state formulation is plotted by the red solid curve,
which overlaps with $\rho_2(-0.9, 0)$. As with the escape time, the
splitting probability is insenstive to the initial position of the
ions, and is dominated by the effects of the potential. 

Next we compare the escape rates $\beta$ in the  two formulations in
Fig. \ref{fig:beta_1well}. The black solid curve shows 
the rate at which an ion at left state $S_2^{(1)}=(-0.9,0)$ escapes
from the left side,  
the black dash-dotted curve plots the rate at which an ion at right
state $S_2^{(2)}=(0,0.9)$ escapes from right side. The left and right
escaping rates at the balanced state $S_2^{(1)}=(-0.45, 0.45)$ in
three-state formulation (\ref{threestate}) are plotted by red solid 
curve and red dash-dotted curve, respectively. 
All escaping rates drop rapidly as $d$ increases. When $d \ge 0.7$,
the potential well is so broad that the two ions are trapped  
in the channel for a long time. In Fig. \ref{fig:gamma_1well}, we plot
the transition rates between states $S_2^{(1)}=(-0.9,0)$ and 
$S_2^{(2)}=(0,0.9)$. Due to the inclined voltage, the transition from 
left state $S_2^{(1)}$ to $S_2^{(2)}$ occurs more often than the other way, 
namely the transition rate $\gamma_2^{(2,1)}$ depicted by solid curve 
is above $\gamma_2^{(1,2)}$ by the dash-dotted curve.
When $d<0.4$, the transition rates are very low ($<10^{-4}$
ns$^{-1}$), so the two states are very distinct. For large $d$, the
observed single state in Fig. \ref{fig:P2} can be 
treated as average of the two distinct states.   

We remark that the mean escape time of the ion from a one-ion channel
$\tau_1(S_1^{(1)}=\{0\})$ is $O(10^5)$ times larger than
$\tau_2(S_2^{(1)})$, so the escape rates $\beta_1^{-(1,1)} \ll
\beta_2^{-(1,1)}$ and   
$\beta_1^{+(1,1)} \ll \beta_2^{+(1,2)}$.  
By \eqref{P2P1P0_formula} we see that for there to be an appreciable
probability of having no ion in the channel we need 
$\tau_1(S_1^{(1)})^{-1} \sim \alpha^{-(1,1)}_0 +
 \alpha^{+(1,1)}_0 = \lambda + \mu$.
In our example, we choose the smallest entry rates to be $\lambda =
\mu = 1$, so the resulting $\widetilde{P}_0^{(1)}$ is extremely
small compared to  $\widetilde{P}_1^{(1)}$ and
$\widetilde{P}_0^{(1)}$, 
and is therefore negligible. 
Thus,  for any entry rates $\lambda=O(1)$, $\mu=O(1)$, only one-ion
occupancy 
(small entry rates) or two-ion occupancy (large entry rates) is
observed most of time (i.e. $\widetilde{P}_1^{(1)} + \widetilde{P}_2^{(1)}
+ \widetilde{P}_2^{(2)} \approx 1$).   
 
\begin{figure}[t]
\centering
\subfigure[]
{\label{fig:P2P1_d}
\includegraphics[width=2.45in]{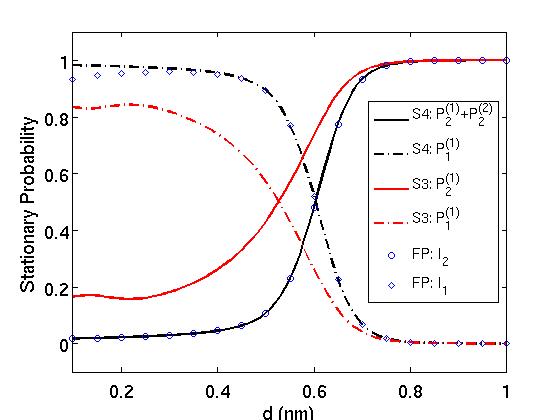}}
\subfigure[]
{\label{fig:flux_1well}
\includegraphics[width=2.45in]{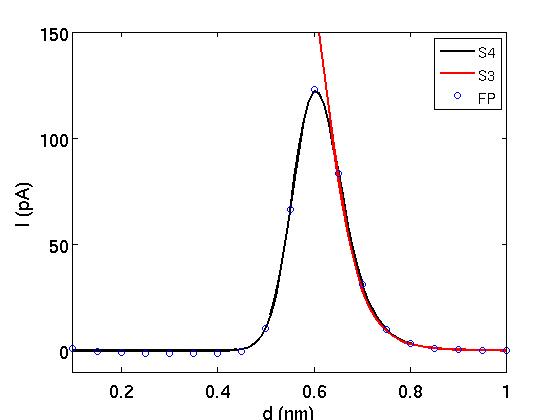}}
\caption{Fixing  $\lambda=\mu=5$ ns$^{-1}$, $U=-0.05$ V nm$^{-1}$. 
{\rm (a)} We plot the stationary probabilities of $2$-ion and $1$-ion by four-state formulation 
$(\ref{fourstate})$
(black curves), three-state formulation $(\ref{threestate})$ (red curves) and
by solving Fokker-Planck equation with $28800$ elements (discrete marker). 
{\rm (b)} We plot the current $I$ vs. $d$ by four-state formulation 
$(\ref{fourstate})$
(black solid curve) and by three-state formulation $(\ref{threestate})$ (red solid curve), which 
are compared with $(f_L+f_R)$ in \eqref{flux} from solution of 
the Fokker-Planck equation (discrete marker). 
}
\end{figure}

In Fig. \ref{fig:P2P1_d}, fixing $\lambda= \mu=5$ ns$^{-1}$, $U=-0.05$
V nm$^{-1}$, we  compare the stationary probabilities of two-ion occupancy
(solid curves) and one-ion occupancy (dash-dotted curve) for various values of
$d$ using (i) the four-state formulation (\ref{fourstate}) (black curves), 
(ii) the three-state formulation (\ref{threestate}) (red curves), and 
(iii) by solving the Fokker-Planck equations using {\em Comsol}  
(discrete markers). When $d$ is large, the stationary probabilities 
obtained from all three methods agree with each other, which confirms
that the single balanced state in the three-state formulation can be
treated as an average of the two distinct 
states (with frequent transitions) in  the four-state
formulation. Because the mean escape time grows from $O(10^{-2})$ ns
to $O(10^2)$ ns with $d$ increasing as shown in Fig. \ref{fig:tau2_d},  
a constant entry rate $\lambda=\mu=5$ ns$^{-1}$ leads to a transition
from initially one-ion  dominant to two-ion 
dominant channel.   

After obtaining the rates and probabilities, we can compare the flux
through the channel from the rate theory and the solution of the
Fokker-Planck equations. 
We integrate the right hand side of \eqref{SSFP2} with respect to
$x_1$ over $[-L, L]$, 
which yields 
\begin{equation}
 \label{integral}
f_{1R} - f_{1L} + (\mu+\lambda) I_0 - (\mu \int_{-L}^{x_+} \widetilde{P}_1 \, d x_1 +\lambda \int_{x_-}^L \widetilde{P}_1 \, d x_1)  - (f_{2R}-f_{2L}) = 0,
\end{equation}
where 
\[ \begin{aligned}
f_{2L} &=  \int_{-L}^L  D \left(
\frac{\partial \widetilde{P}_{2}}{\partial x_1}  + 
\widetilde{P}_{2}  \frac{\partial \Phi_2}{\partial x_1}\right)(-L,x_1) \, d x_1 
= \int_{-L}^L D \frac{\partial \widetilde{P}_{2}}{\partial x_1}(-L,x_1) \, d x_1 ,\\
f_{2R} &=  \int_{-L}^L D \left( 
\frac{\partial \widetilde{P}_{2}}{\partial x_2}  + 
\widetilde{P}_{2}  \frac{\partial \Phi_2}{\partial x_2}\right)(x_1,L) \, d x_1
= \int_{-L}^L D \frac{\partial \widetilde{P}_{2}}{\partial x_2}(x_1,L) \, d x_1, \\
f_{1L} &=   D \left( \frac{d \widetilde{P}_1}{d x_1} + \widetilde{P}_1 \frac{d \Phi_1}{d x_1} \right)(-L) 
= D \, \frac{d \widetilde{P}_1}{d x_1}(-L), \\
f_{1R} &=   D \left( \frac{d \widetilde{P}_1}{d x_1} + \widetilde{P}_1 \frac{d \Phi_1}{d x_1} \right)(L) 
= D \, \frac{d \widetilde{P}_1}{d x_1}(L).  
\end{aligned}
\]
In Table \ref{tabletransitions}, we show that the eight transitions
connected to state $S^{(1)}_1$ in Fig. \ref{fig:ill1} correspond 
one by one to the eight terms in \eqref{integral}. 

\begin{table}[t]
\begin{center}
\begin{tabular}{|c|c|c|}
\hline
Transitions by escaping  & Flux from rate theory & Flux from Fokker-Planck \\
\hline
$S^{(1)}_2 \xrightarrow{\beta^{-(1,1)}_2} S^{(1)}_1$ &
$\beta^{-(1,1)}_2 \widetilde{P}^{(1)}_2$ &
$f_{2L}$ \\
$S^{(2)}_2 \xrightarrow{\beta^{+(1,2)}_2} S^{(1)}_1$ & 
$\beta^{+(1,2)}_2  \widetilde{P}^{(2)}_2$ &
$-f_{2R}$ \\
$S^{(1)}_1 \xrightarrow{\beta^{+(1,1)}_1} S^{(1)}_0$  &  
$\beta^{+(1,1)}_1 \widetilde{P}^{(1)}_1$ &
$f_{1L}$ \\
$S^{(1)}_1 \xrightarrow{\beta^{-(1,1)}_1} S^{(1)}_0$ & 
$\beta^{-(1,1)}_2  \widetilde{P}^{(1)}_1$ &
$- f_{1R}$ \\  
$S^{(1)}_1 \xrightarrow{\alpha^{-(1,1)}_1}S^{(1)}_2$ &  
$\mu \widetilde{P}^{(1)}_1$ &
$\mu I_1$ \\
$S^{(1)}_1 \xrightarrow{\alpha^{+(2,1)}_1}  S^{(2)}_2$ &
$ \lambda \widetilde{P}^{(1)}_1$ &
$\lambda  I_1$ \\
$S^{(1)}_0 \xrightarrow{\alpha^{-(1,1)}_0}  S^{(1)}_1$  &  
$\mu \widetilde{P}^{(1)}_0$ &
$\mu I_0$ \\
$S^{(1)}_0 \xrightarrow{\alpha^{+(1,1)}_0}  S^{(1)}_1$ & 
$ \lambda \widetilde{P}^{(1)}_0$ & 
$\lambda  I_0$ \\
\hline 
\end{tabular}
\end{center}
\caption{Transitions computer by the four-state model (\ref{fourstate}) and
the hierarchical Fokker-Planck equation.}
\label{tabletransitions}
\end{table}

The left to right flux across the left boundary of the channel is generated by 
introducing new ions $\mu I_0 + \mu \int_{-L}^{x_+} \widetilde{P}_1 \, d x_1 $, and the right to 
left flux across the 
left boundary of the channel is generated by ions leaving the left boundary $f_{1L} +  f_{2L}$.
Similarly we have the left to right flux across the right boundary of the channel is generated by
ions leaving the right boundary $ - f_{1R} - f_{2R}$, and  the right to left flux across the 
right boundary of the channel is generated by introducing new ions
$\lambda I_0 + \lambda \int_{x_-}^L \widetilde{P}_1 \, d x_1$. Thus
the overall fluxes are
\begin{equation}
\label{flux}
f_L =  \mu I_0 + \mu \int_{-L}^{x_+} \widetilde{P}_1 \, d x_1  -  f_{1L} -  f_{2L}  , \quad
f_R = -  f_{1R} -  f_{2R} - \lambda I_0 - \lambda \int_{x_-}^L \widetilde{P}_1 \, d x_1.  
\end{equation}

In Fig. \ref{fig:flux_1well}, we fix $ \lambda=\, \mu=5$ ns$^{-1}$, 
$U=-0.05$ V nm$^{-1}$, and
plot the current $I$ vs. $d$ from the flux $(f_L+f_R)$ in \eqref{flux} 
by discrete open diamands,
obtained by solving \eqref{FP2}. In comparison, the black solid curve
depicts the current obtained  by applying the four-state rate theory
(\ref{fourstate}), and the red solid
curve depicts the current by the three-state rate theory (\ref{threestate}).  
We  see that for a broad potential well with  $0.6 < d < 1$,
all three methods reach a good agreement: 
the four-state works for large $d$, even though
the stationary probability distribution in Fig. \ref{fig:P2P1_d} shows
there should be three states for large $d$. This is because of our use
of escape times and splitting probabilities to determine the
transition rates in the model, rather than by trying to estimate
hopping rates directly.
However,
with small $d$, the  three-state model is not an accurate description of the
Markov process, and the flux  quickly becomes inaccurate.  

An important observation from Fig. \ref{fig:flux_1well} is that a maximal
current is achieved around $d=0.6$ nm, which means there exists an
optimal 
shape of the potential well to conduct ions, even if the depth of the well 
remains the same. This result agrees with the argument 
in \cite{Abad:2009:NRT} for a single 
ion channel with piecewise linear potential energy. 

We may explain the existence of an optimal flux by looking at the
stationary probabilities  $\widetilde{P}_2$ and $\widetilde{P}_1$ as
a function of $d$ in Fig. \ref{fig:P2P1_d}. 
When $d$ is small, the potential well is very narrow and steep near 
the binding site, but relatively flat near the end of channel, so any new ion
introduced would escape very quickly from the same end by diffusion, 
leaving the old ion in the channel. For example, at $d=0.1$ nm, both 
escaping rates $\beta_2\left(S_2^{-(1,1)}\right)$ and 
$\beta_2\left(S_2^{+(1,2)}\right)$
are over $500$ ns$^{-1}$.  Thus the channel has only 
one ion ($\widetilde{P}_1 > 0.8$) most of the time, obviously the process of 
ion entering and leaving from the same end 
does not generate any through flux. On the other hand, when $d$ is 
large, the potential well is very broad and flat near the binding site, 
two ions can hardly escape from the channel, as shown by the large 
mean escape time in Fig. \ref{fig:tau2_d}, once a new ion is introduced 
to the channel,
it quickly moves towards the center, and settles into  a balanced
state in the well with the other ion; thus the $2$-ion state dominates
($\widetilde{P}_2 
> 0.9$).  
In this case the flux is small because two ions are trapped in the
channel for a long time. 

When neither $2$-ion or $1$-ion occupancy dominates in the channel,
so that there are adequate transitions between the $2$-ion distinct states and
frequent escapes from the $2$-ion to the $1$-ion state, a large flux is
generated. This explains heuristically why an intermediate potential well has an
optimal geometry. 

\begin{figure}[t]
\centering
\subfigure[]
{ \label{fig:Prob_d_vary_mu}
 \includegraphics[width=2.45in]{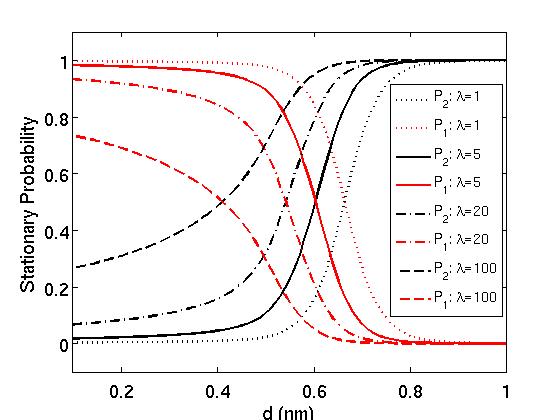}}
\subfigure[]
{ \label{fig:I_d_vary_mu}
 \includegraphics[width=2.45in]{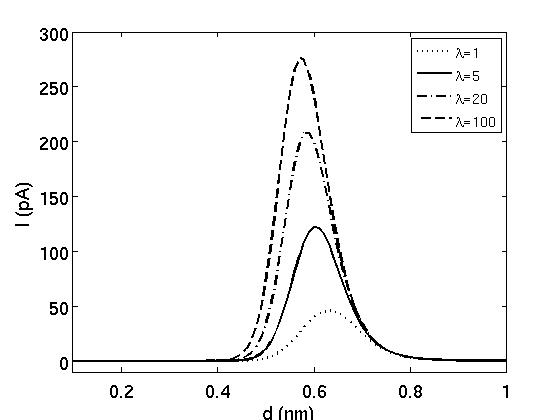}}
\caption{Fixing $U=-0.05$ V nm$^{-1}$.  
{\rm (a)} We plot stationary probability $\widetilde{P}_2$ (black) and $\widetilde{P}_1$ (red) 
obtained from four-state Markov Chain formulation $(\ref{fourstate})$ for $\lambda=1, 5, 20, 100$ ns$^{-1}$. 
The intersection points of each pair of curves at which $\widetilde{P}_2 = \widetilde{P}_1 \approx 0.5$ 
for $\lambda=1, 5, 20, 100$ are at $d \approx 0.665, 0.6, 0.54, 0.41$ nm, respectively.
{\rm (b)} We plot current obtained from four-state Markov Chain formulation for $\mu=1, 5, 20, 100$ ns$^{-1}$. 
The critical values of $d$ at which optimal flux is achieved are respectively 
$d \approx 0.63, 0.6, 0.59, 0.57$ nm.
}
\end{figure}

Finally we investigate how the entry rates affect the optimal flux for
the family of potential wells  in
Fig. \ref{fig:pot_group_1well:a} using  the four-state Markov chain
formulation (\ref{fourstate}).  
Recall that the escape rates $\beta_2^{-(1,1)}$,  $\beta_2^{+(1,2)}$ and transition 
rates $\gamma_2^{(1,1)}$, $\gamma_2^{(1,2)}$ are determined by the
potential through mean escape time and left splitting probability, and
thus are independent of the entry rates. We fix the applied field
$U=-0.05$ V nm$^{-1}$ 
and plot in Fig. \ref{fig:Prob_d_vary_mu} the stationary probabilities
$\widetilde{P}_2$ (black) and $\widetilde{P}_1$ (red) for $\lambda=1,
5, 20, 100$ ns$^{-1}$. 
The intersection points of each pair of curves at which
$\widetilde{P}_2 = \widetilde{P}_1 \approx 0.5$ for $\lambda=1, 5, 20,
100$ are at $d \approx 0.665, 0.6, 0.54, 0.41$ respectively. This
illustrates the fact that when the potential well is narrow and steep
at the binding site ($d$ small), the escaping rates $\beta_2^{-(1,1)}$
and  $\beta_2^{+(1,2)}$ are  large (shown in
Fig. \ref{fig:beta_1well}), so the  entry rates have to increase in
order to have equal probabilities of $2$-ion and $1$-ion occupancy.  

In Fig. \ref{fig:I_d_vary_mu}, we plot the current $I(d)$ for
$\lambda=1, 5, 20, 100$ ns$^{-1}$. As expected, the current increases
as the entry rates increases, but we also find that the value of $d$
at which the current is optimised shifts;
the critical values of $d$ at which optimal flux is achieved are
respectively $d \approx 0.63, 0.6, 0.59, 0.57$. Thus the optimal
value of $d$ slightly decreases as the entry rates increases, which
shows that the larger 
escaping rates of tighter potentials require larger entry rates to
optimize the flux. 

\section{Conclusion} \label{sec:conclusion}

We have presented a set of hierarchical Fokker-Planck equations
describing ion permeation in multi-ion channels, and reduced these
systematically to
a discrete rate theory. The basis of the reduction is the fact  that
many channels have internal binding sites at which ions sit, 
so that ions transport by hopping between sites on a slow time scale
while oscillating
in the binding sites on a fast time scale. Since the fast oscillation is
not key in determining the conduction rate, we can
reduce the continuous dynamics to the slow transition between the
discrete states, 
and thus provide an efficient way to calculate  the current through
the channel. 
A key component of our reduction was the use of exit times and
splitting probabilities to determine the discrete hopping rates,
rather than trying to estimate these directly using Kramer's theory for
example. This means that the predictions of the discrete model are
accurate even when the internal states are not so well defined.

In contrast to traditional Eyring rate theory \cite{Eyring:1935:ACA} and
the recent study of a one-ion channel in \cite{Abad:2009:NRT}, we have
developed a general theory
for multi-ion channels, and have shown an intricate coupling between
transition rates, mean escape time and splitting probability, due to
the complexity of
the resulting system of Markovian states. The theory is illustrated by
a two-ion channel, which is the most accessible example that
includes the multi-ion complexity. We have investigated how
conductivity of the channel depends on the diffusion coefficient,
potential energy landscape,
and the ion entry rate. By
varying the geometry of the external potential while keeping the 
depth fixed, we observed that
when the potential well is narrow and steep at the binding site, the
$1$-ion state dominates, but when it is not the $2$-ion state
dominates.
In between there is an optimal geometry which maximizes the ion flux by
negotiating between these two extremes  and allowing
frequent transitions between the $1$-ion and $2$-ion states.

\vskip 5mm

\noindent
{\bf Acknowledgements.}
This publication was based on work supported by Award
No KUK-C1-013-04, made by King Abdullah University of Science
and Technology (KAUST). The research leading to these results has received
funding from the {\it European Research Council} under the
{\it European Community's} Seventh Framework Programme
({\it FP7/2007-2013})/ ERC grant agreement No. 239870.
Radek Erban would also like to thank
Somerville College, University of Oxford, for a Fulford Junior
Research Fellowship; Brasenose College, University of Oxford, for a
Nicholas Kurti Junior Fellowship; the Royal Society for a University
Research Fellowship; and the Leverhulme Trust for a Philip
Leverhulme Prize.

\newpage

\bibliographystyle{siam}

\end{document}